\documentclass[12pt]{article}

\usepackage[dvips]{graphicx}
 \font\gotb eufm10 scaled \magstep1
\newcommand{\bb}{\bibitem}
\newcommand{\cc}{\cite}
\newcommand{\vp}{\varphi}
\newcommand{\vt}{\vartheta}
\newcommand{\sss}{\sigma}
\newcommand{\al}{\alpha}
\newcommand{\Om}{\Omega}
\newcommand{\om}{\omega}
\newcommand{\lt}{\left}
\newcommand{\rt}{\right}

\newcommand{\ra}{\rangle}
\newcommand{\lll}{\lambda}
\newcommand{\F}{{\cal F}}

\newcommand{\D}{\hat D}

\newcommand{\s}{\hat S}
\newcommand{\K}{\hat K}
\newcommand{\I}{\hat I}

\newcommand{\A}{\hat A}
\newcommand{\B}{\hat B}
\newcommand{\U}{\hat U}
\newcommand{\V}{\hat V}
\newcommand{\W}{\hat W}
\newcommand{\p}{\hat p}

\newcommand{\AAA}{\hbox{\gotb A}}

\newcommand{\HHH}{\hbox{\gotb H}}
\newcommand{\QQ}{\hbox{\gotb Q}}
\newcommand {\QQQ}{\hbox {\gotb Q}_{\xi}}
\newcommand {\qqq}{\hbox {\gotb Q}_{\xi'}}
\newcommand {\vx}{\vp_{\xi}}

\newcommand{\OO}{{\cal O}}

\newcommand{\bea}{\begin{eqnarray} \label}
\newcommand{\eeq}{\end{equation}}
\newcommand{\beq}{\begin{equation} \label}
\newcommand{\eea}{\end{eqnarray}}
\newcommand{\nn}{\\ \nonumber}
\newcommand{\rr}[1]{(\ref{#1})}

\newcommand{\post}{{\sc Postulate}}

 \author{\it D.A.Slavnov }

\title{The Locality Problem in Quantum Measurements
\thanks{Physics of Particles and Nuclei, 2007, Vol. 41, No. 1, pp. 149-173.} }

   \date{}

\begin{document}
\maketitle
\begin{center}

\begin{center} {\it  Department of Physics, Moscow State
University,\\  Moscow 119899, Russia. E-mail:
slavnov@goa.bog.msu.ru }
 \end{center}

\end{center}

 \begin{abstract}

The locality problem of quantum measurements is considered in the
framework of the algebraic approach. It is shown that contrary to
the currently widespread opinion one can reconcile the
mathematical formalism of the quantum theory with the assumption
of the existence of a local physical reality determining the
results of local measurements. The key quantum experiments:
double-slit experiment on electron scattering, Wheeler's
delayed-choice experiment, the Einstein-Podolsky-Rosen paradox,
and quantum teleportation are discussed from the locality-problem
point of view. A clear physical interpretation for these
experiments, which does not contradict the classical ideas, is
given.
\end{abstract}

\

\section{INTRODUCTION}

The locality problem is one of the main problems in the entire
quantum theory. It attracted especially close attention during the
construction of the quantum field theory (see, e.g.,
\cc{bog,wigh,araki,haag}), where the locality axiom plays a
central role. This axiom has different formulations; however,
without going into mathematical subtleties, it can be reduced to
the following: boson fields must commute at space-like separated
points, while fermion fields must anti-commute. If the dynamics of
a quantum system is described by a Lagrangian, it is additionally
required that this Lagrangian must be a local function of the
fields.

The following argument is often used as a physical justification
of the locality axiom. The results of the measurement in a bounded
domain of a Minkowski space (a local measurement) are determined
by boson-field values and by bilinear combinations of fermion
fields in this domain.

Such locality requirement is purely mathematical in its nature. It
can be formulated only in the framework of a particular
mathematical formalism, and it is a part of that formalism. In a
general discussion of locality it is desirable to proceed from
requirements that can be formulated in physics terminology and
that can be checked in the experiment directly. Such formulation
is fairly obvious.

If two bounded domains $\OO_1$ and $\OO_2$  of the Minkowski space
are space-like separated, then the results of measurements in the
domain $\OO_1$ do not depend on any manipulations in the  $\OO_2$.

At this point one has to make the following warning. The above
requirement does not imply that the measurement results in the
domains  $\OO_1$ and $\OO_2$ may not be related to each other.
This of course is not true. Such measurement results may have a
common cause, and therefore, a correlation between them is quite
possible.

Practically no one argues with the above formulation. However, the
situation changes radically when we try to supplement the above
requirement with the following one. There exists a certain
physical reality, which determines the results of a local
measurement.

Many people object to such an extension of the locality
requirement. The arguments on this matter began a long time ago.
One can recall the famous debates between Einstein and Bohr.
Einstein (see \cc{epr,ein1,ein2}) was in favor of the above
extension, while Bohr (see \cc{bohr1,bohr2}) was against it.

Later on, the majority's opinion within the physics scientific
community leaned towards the Bohr side. The results of many modern
experiments related to this problem are currently considered as
proof that the physical reality mentioned above does not exist.
Or, at least, the assumption of the existence of such a reality
contradicts the accepted mathematical formalism of the quantum
theory.

However, if we abandon the extension formulated above, we almost
completely lose the physical foundation behind the locality axiom
accepted in the quantum field theory. This rejection would force
us to assume that neither local fields, nor their combinations
describe a local reality (because it does not exist). Then, it is
not clear why these combinations must commute in space-like
separated domains.

Thus we have a deadlock situation. The assumption of the existence
of a local physical reality contradicts the mathematical formalism
of the quantum theory. At the same time, the rejection of this
assumption denies the physical foundation one of the main axioms
in the mathematical formalism of the quantum field theory. Of
course, one can abandon attempts of any physical justification of
the locality axiom. This would affect neither the logical nor
mathematical structure of the quantum theory. However, from the
physical point of view this way out of the uncomfortable situation
is extremely undesirable.

In the present paper we will attempt to demonstrate that the
mathematical formalism accepted within the quantum theory is quite
compatible with the assumption of the existence of physical
reality determining the results of local measurements. The
often-produced incompatibility proofs have the following two
flaws. First, these proofs often point out toward a contradiction
between the experimental data and certain mathematical
assumptions, which are used in the construction of mathematical
formalism. However, the questions of physical validity of these
assumptions and their necessity are usually not discussed. Second,
the interpretation given to the obtained experimental data is far
from being always adequate.

The so-called de Broglie waves  \cc{dbr} can be considered as one
of the most striking examples of inadequate interpretation. In the
beginning of practically any textbook on quantum mechanics it is
said that a de Broglie wave with the wavelength
 \beq{1}
\lll=\frac{2\pi\hbar}{k}.
  \eeq
is associated with any quantum particle having the momentum $k$.
The results of electron diffraction experiments  \cc{dg,tr}, or
the later results on the electron interference  \cc{mss}. are
mentioned as examples supporting the above statement. In agreement
with  \rr{1} a clear interference pattern was observed in the
latter experiment.

 \begin{figure}[h]
 \begin{center}

  \includegraphics{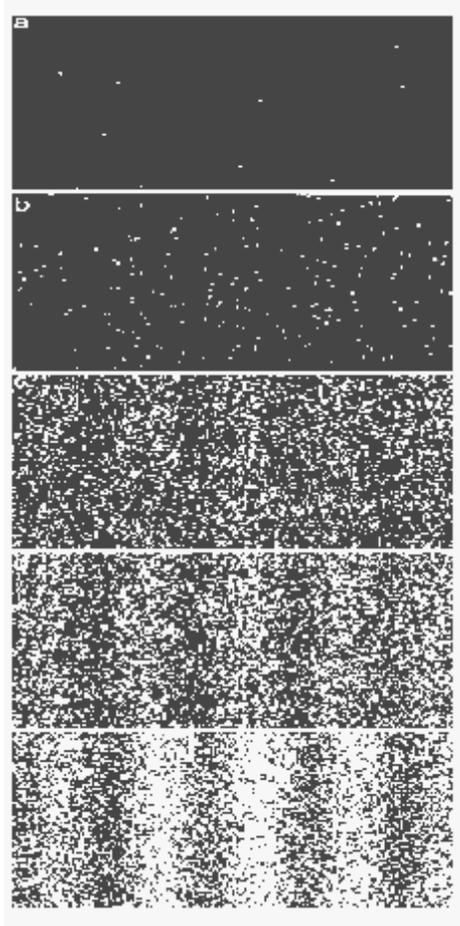}

   \caption{Interference pattern in electron scattering.}
\end{center}
\end{figure}

Equation \rr{1} became the basis of subsequent assertions, that
the distinctive feature of quantum particles is the presence of
both corpuscular and wave properties. These assertions seem to be
quite well supported experimentally. Nevertheless, we would like
to examine if this is indeed the case.

Let us turn to the results of more recent experiments performed by
Tonomura  \cc{ton1,ton2}. These experiments investigated electron
beam scattering by a biprism, which by its physical properties is
analogous to a double-slit screen. The beam intensity was so low,
that on average there was less than one electron in the
experimental apparatus at any single moment. This allowed one to
neglect the influence of inter-electron interaction on the results
of the experiment. Moreover, it was possible to register the
results of passage of a small number of electrons in this
experiment.

The experimental results are shown in Fig. 1, reprinted from the
review \cc{ton2}. The individual photographs correspond to
different exposure times. The photograph (a) registered traces of
10 electrons, the photograph (b) 200 electrons, the photograph (c)
6000 electrons, the photograph (d) 40000 electrons, and the
photograph (e) 140000 electrons.

We see that when only a small number of electrons are registered
(the photographs (a) and (b)) the interference pattern is not
showing through. Such a pattern appears only after a very large
number of electrons were registered (the photographs (d) and (e)).
If we try to determine the electron wavelength with a help of the
photographs (a) and (b), we do not obtain anything similar to
\rr{1}.

These results speak in favor of the fact that wave properties are
not revealed by a single electron. They become apparent only in a
specially prepared ensemble of electrons. In the considered case,
all electrons had approximately the same momentum.

Now we can look at Eq. \rr{1} in a new light. We can assume that
it is relevant not to a single electron, but to a specially
prepared ensemble of electrons. At this point it is appropriate to
say that Blokhintsev \cc{blo1,blo2,blo3,blo4} was a passionate
supporter of the view that most of the quantum mechanics formulas
are relevant not to isolated quantum particles, but to their
ensembles. This novel point of view on the connection between
momentum and spatial characteristics of quantum objects might
prove to be very important for discussions of the locality
properties. The localization domain of a single object is one
thing, but the localization domain of an ensemble of these objects
is an entirely different story. Later on we will come back to
discussion of this question.

The commonly used in textbooks formulation of the mathematical
formalism of the quantum theory, with wave functions or state
vectors as the basic elements, is not ideal for discussions of the
locality problem, because these objects themselves are obviously
nonlocal. The so-called algebraic approach  \cc{emch,hor,blot} is
much better suited for these purposes. Unlike the traditional

approach, the Hilbert's space of state vectors is no longer a
primary object of the theory within the algebraic approach, and
observables are no longer defined as operators in the Hilbert
space.

Observables, more specifically, local observables are considered
as the primary elements of the theory. Heuristically, an
observable is defined as such an attribute of the investigated
physical system for which one can obtain some numerical value with
the help of a certain measuring procedure. Accordingly, for local
observables one can obtain numerical values with the help of local
measurements. Usually it is assumed that some unit system is
chosen, and therefore one can consider all observables as
dimensionless.

Initially the observables are not related to operators in a
Hilbert space at all. The Hilbert space itself is constructed with
the help of observables as some secondary object. After that a
connection between the observables and the operators in this space
is established.

We will conduct the subsequent examination in the framework of a
special version of the algebraic approach previously described in
detail in the paper \cc{slav1}. A more concise description can be
found in the papers  \cc{slav2,slav3}. In the papers
\cc{slav1,slav2} the theory was constructed using an inductive
approach, where physical laws are noticed first, and then they are
formulated as mathematical axioms. In the present paper we will
use a deductive approach, where the main axioms are formulated
from the very beginning. We refer the readers wishing to get
acquainted with physical justifications of these axioms to the
paper  \cc{slav1}. The main definitions and statements from the
theory of algebras can be found in the same paper.

\section{THE MAIN ASPECTS OF THE ALGEBRAIC\\ APPROACH}

We begin from stating the basic properties of observables. The
main property is the following one. The observables can be
multiplied by real numbers, added to each other, and multiplied by
one another. This property is formulated as the following
postulate.

\

\post{} 1. The observables  $\A$ of a physical system are
Hermitian elements of some $C^*$-algebra.

\

Postulate 1 (and all the subsequent ones) is valid for classical
systems as well. We remind the reader that algebra is a set of
elements, which, firstly, is a linear space; secondly, a
multiplication operation is defined for pairs of elements from
that set. That is, for any two elements
 $\U,\V\in\AAA$ a third element  $\W\in\AAA$ is
assigned, and the following notation is used  $\W=\U\V$. This
assignment satisfies a number of properties, which are standard
for multiplication operations. An algebra
 $\AAA$ is a $C^*$-algebra (see, e.g.,~\cc{dix}), if a conjugation
 operation (involution) $\U\to\U^*$ $(\U\in\AAA,\quad
\U^*\in\AAA)$ is defined on $\AAA$, and the norm of any element
$\U$ satisfies the condition  $\|\U\U^*\|=\|\U\|^2$. The
justification why the algebra $ \AAA$ must be a  $C^*$ algebra can
be found in \cc{slav1}. An element  $\U$ is called Hermitian, if
$\U^*=\U$. The set of observables will be denoted  $\AAA_+$
($\AAA_+\subset\AAA$). In classical systems all observables are
compatible with each other (can be measured simultaneously). In a
quantum system they can be either compatible or incompatible.

\

\post{} 2. The set of compatible with each other observables is a
maximal real associative commutative subalgebra~$\QQQ$ of the
algebra~\AAA \quad ($\QQQ\subset\AAA_+$).

\

The index~$\xi$, which runs through the set  $\Xi$, distinguishes
one such subalgebra from another. For a classical system the set
$\Xi$ contains just a single element, for a quantum system  $\Xi$
contains infinitely many elements.

Of course, in the framework of the traditional approach to the
quantum theory, the bounded observables satisfy Postulates 1 and
2. However, there is an additional requirement within the
traditional approach, that the observables are described by
self-adjoint operators defined in a Hilbert space. This
requirement allows one to construct a very efficient mathematical
formalism; however, it does not have an intuitive physical
foundation. Moreover, the necessity of this additional requirement
is not quite clear.

The set of observables  $\AAA_+$ can be considered as a
mathematical model of a quantum system. Accordingly, the subset
$\QQQ$ can be considered as observables of some classical
subsystem. This subsystem is open, because the quantum-system's
degrees of freedom corresponding to observables from different
subsets  $\QQQ$ can interact with each other.

Moreover, these classical subsystems may not have their own
dynamics, because the generalized coordinates and momenta
corresponding to the same degree of freedom, may belong to
different subsets of  $\QQQ$. Therefore, the traditional approach
for defining the state as a point of a phase space is not suitable
for such subsystems. But, specifying a point in the phase space is
equivalent to setting initial conditions for the equations of
motion. This allows one to fix the values of all observables of
the considered system. However, one can avoid using equations of
motion and the initial condition, and fix the values of all
observables directly. Such an approach is suitable for open
systems as well.

Measuring the sum of observables in any concrete classical system
yields the sum of the values of the individual observables, and
measuring the product of observables yields the product of their
individual values. In other words, specifying the values of all
observables is equivalent to specifying some homomorphic map of
the algebra of observables into the set of real numbers. For
commutative associative algebra, such a map is called a character
(see, e.g.,~\cc{dix}). Therefore we accept the following
postulate.

\

\post{} 3. The state of a classical subsystem, whose observables
are elements of a subalgebra  $\QQQ$, is described by a character
of this subalgebra.

\

This definition of the state of a classical subsystem has an
important advantage, that it can be generalized to the quantum
case. Each quantum observable belonging to  $\AAA_+$,
simultaneously belongs to some subalgebra  $\QQQ$. This allows one
to consider a quantum system as a family of classical subsystems.
If we knew the states of all these subsystems, we could have
predicted the result of measuring any observable of the quantum
system. This gives us grounds for accepting the following
postulate.

\

\post{} 4. The result of measuring any observable of a quantum
system is determined by its elementary state   $\vp$.

\

Here, $\vp$ is a family  $\vp=[\vx]$ of characters  $\vx$ of all
subalgebras $\QQQ$. Each subalgebra  $\QQQ$ in the family is
represented by a single character.

At first it may seem that the last postulate contradicts the fact
that one cannot predict the measurement results for all
observables of a quantum system. However, there is no
contradiction here. The point is that we can measure
simultaneously (that is in a compatible way) only compatible
observables. These observables belong to a certain subalgebra
$\QQQ$. Lets say for instance they belong to the subalgebra with
the index  $\xi=\eta$. Then, from the complete set $[\vx]$ we can
specify only one character  $\vp_{\eta}$. Endowed with such
information we can predict only the measurement results for
observables belonging to  $\QQ_{\eta}$.

We will not be able to say anything certain about the values of
other observables. Additional measurements, if they are not
compatible with the previous ones, will not improve the situation.
They will produce new information about the quantum system;
however, simultaneously the additional measurements will disturb
the state of our system and will make the information obtained
earlier worthless.

Figuratively speaking, an elementary state is a holographic image
of the system under investigation. Using classical measuring
devices we can look at it from one side only, and, hence, obtain a
two-dimensional image. Moreover, the measurement will disturb the
system and will change its original holographic image. Therefore,
if later we will look at the system from another side, we will see
a two-dimensional projection of the new holographic image. Thus,
we will never be able to see the entire holographic image.

Using the notion of an elementary state, one can see the Everett
idea  \cc{ev} on the existence of many parallel worlds in a new
light. According to his original idea quantum systems
simultaneously occupy many parallel worlds, while a classical
observer lives in only one of these worlds chosen accidentally.
Therefore, he sees only the version of the quantum system
represented in his world. In this form the Everett idea looks
implausible.

On the contrary, the idea that an elementary state is analogous to
a holographic image looks quite plausible. On the other hand, the
notion of an elementary state leads to roughly the same
consequences as the idea of the existence of many worlds. However,
there is an important difference. In Everett's picture the
randomness inherent to quantum measurements is connected with an
accidental occupation of one or another world by the classical
observer. In the language of probability theory, in this case, we
are dealing with an ensemble of observers. When we describe a
quantum system with the help of an elementary state, we obtain an
incomplete description of the system. Therefore, in reality we
describe not an individual system, but certain characteristics
common for an entire ensemble of quantum systems.

In connection with the above it is useful to introduce the notion
of $\vp_{\eta}$-equivalence. Two elementary states $\vp=[\vx]$ è
$\vp'=[\vx']$ will be called  $\vp_{\eta}$-equivalent, if
$\vp_{\eta}=\vp'_{\eta}$. The relations between the remaining
characters $\vx$ and  $\vx'$ can be arbitrary. The class of
$\vp_{\eta}$-equivalent elementary states will be denoted
$\{\vp\}_{\vp_{\eta}}$. The most that one can possibly learn about
an elementary state $\vp$  is that it belongs to some equivalence
class $\vp\in\{\vp\}_{\vp_{\eta}}$.

There is one more obstacle preventing unambiguous predictions of
measurement results. One and the same observable
 $\A$ may belong simultaneously to several subalgebras
 $\QQQ$: $\A \in \QQQ\cap\qqq$
$(\xi\neq\xi')$. Therefore, it is not clear which of the
functionals (characters)  $\vx$ or  $\vp_{\xi'}$ will describe the
results of a particular measurement.

At first it may seem that this additional ambiguity can be easily
eliminated with the help of the additional condition
  \beq{2}
 \vx(\A)=\vp_{\xi'}(\A), \mbox{ åñëè }\A \in \QQQ\cap\qqq.
\eeq
 However, this condition leads to numerous contradictions.
 Additional details can be found in the paper  \cc{slav1}.
 On the other hand, as was shown in the same paper \cc{slav1},
 the condition  \rr{2} is not a necessary one. Indeed,
 the measurement result may depend not only on the system
 under investigation, but on the characteristics of the
 measuring device as well. From the observer's point of
 view such dependence is extremely objectionable, and
 experimentalists try to minimize it as much as possible.

We have come to think that measurement results are virtually
independent of the characteristics of "good" measuring devices.
However, for this to be true all the devices used for measuring
the observable of interest must at least be calibrated in a
consistent way. It was shown in \cc{slav1} that the existence of
incompatible measurements in the quantum case makes such
calibration far from being always possible. In particular, if we
assign a certain type of measuring devices ($\xi$-type) to every
subalgebra  $\QQQ$, then, as it turns out, the devices of
different types cannot be calibrated consistently. Therefore, one
cannot get rid of a possible dependence of the measurement results
on the device type (or, on the index  $\xi$).

The above assertion does not exclude that for some elementary
states  $\vp$ Eq. \rr{2} will be valid for all  $\QQQ$, $\qqq$,
containing the observable  $\A$. In this case we shall say that
the elementary state $\vp$ is stable with respect to the
observable
 $\A$.

Let us go back to the discussion of elementary states and of their
equivalence classes. Measurements allow one to establish that the
elementary state  $\vp$ of the system under investigation belongs
to some equivalence class $\vp\in\{\vp\}_{\vp_{\eta}}$.
Thereafter, we can make the following predictions. Measuring
devices of the  $\eta$-type will yield the value
$A=\vp_{\eta}(\A)$  for the observable $\A\in\QQ_{\eta}$. From now
on the measurement result is denoted by the same symbol as the
observable itself, but without the "hat." If the elementary state
$\vp$ is stable with respect to the observables $\A\in\QQ_{\eta}$,
then the same result will be obtained by using measuring devices
of any type  $\xi$. Of course, the type  $\xi$ must be such that
$\A\in\QQQ$, otherwise the device is simply not suitable for
measuring the values of this observable. One cannot say anything
definite about measurement results for observables
$\A\notin\QQ_{\eta}$, because we will obtain different values for
different elementary states $\vp\in\{\vp\}_{\vp_{\eta}}$,  .

Within the standard mathematical formalism of quantum mechanics
all the physical properties mentioned above are exhibited by
quantum states specified by particular values of a complete set of
commuting observables. This allows one to state the following
definition of a quantum state within the proposed approach.

\

{\sc Definition}. A quantum state  $\Psi_{\vp_{\eta}}$ is the
class  $\{\vp\}_{\vp_{\eta}}$ \quad $\vp_{\eta}$- equivalent
elementary states, which are stable with respect to the
observables $\A\in\QQ_{\eta}$.

\

It is usually assumed that a quantum state  $\Psi_{\vp_{\eta}}$
appears as a result of measuring the observables
$\A\in\QQ_{\eta}$, where a specified value is registered for each
of the observables  $\A$. Of course, this is not always true, at
least, because some particles of the investigated system can be
absorbed by the device in the measuring process. In order for a
measurement to be simultaneously a preparation of a quantum state,
it must be reproducible. If repeated measurements of an observable
$\A$ give identical results, we shall mean the measurements
reproducible. Note that the repeated measurements are not
necessarily performed by measuring devices of the same type.

The paper  \cc{slav1} shows that quantum states defined above are
always pure. Within the standard mathematical formalism of quantum
mechanics pure states are defined as vectors  $|\Phi\ra$ of some
Hilbert space  $\HHH$. These vectors are used for calculating the
average values of observables in the corresponding quantum states.
This definition works very well for applied purposes; however, it
does not have an intuitively clear physical interpretation. Within
the approach proposed in the present paper the average value of an
observable is connected in a natural way with the probability
distribution of the elementary states $\vp$ within the equivalence
class  $\vp\in\{\vp\}_{\eta}$.

One has to bear in mind that the elementary states satisfy the
standard properties of elementary events from the classical
Kolmogorov probability theory  \cc{kol}. Namely, each random
experiment results in one and only one elementary event. Different
elementary events are mutually exclusive. Note that the standard
approach to quantum mechanics does not have such an ingredient.
This became an insurmountable obstacle for application of the
classical probability theory to quantum mechanics. Such an
obstacle is absent within the approach used here. Therefore, there
is no need for creating some artificial quantum probability
theory. Instead one can use the well-developed formalism of the
classical probability theory. Therefore, the following postulate
appears to be fairly natural.

\

\post{} 5. The equivalence class  $\{\vp\}_{\vp_{\eta}}$
corresponding to the quantum state $\Psi_{\vp_{\eta}}$ can be
equipped with the structure of a probability space  $(\Om,\F, P)$.

\

The probability space  $(\Om,\F, P)$ is the main notion of the
Kolmogorov probability theory. Let us remind the reader the
meaning of its components.  $\Om$ is the set (space) of elementary
events. In our case the equivalence class $\{\vp\}_{\vp_{\eta}}$
plays the role of $\Om$, while the elementary states   $\vp$ play
the role of elementary events.

On top of the elementary events the so-called probability events
$F$ also used in the Kolmogorov probability theory. Frequently
they are called simply events. Any event $F$ is a subset of the
set  $\Om$. It is assumed that an event  $F$ has happened in a
random experiment, if one of the elementary events from $F$ has
been realized in the experiment.  $\F$ is a Boolean $\sss$-algebra
of the subsets  $F$. That is,  $\F$ is endowed with the following
three algebraic operations: the union of subsets  $F$, the
intersection of these subsets, as well as the complement of each
subset to the entire set $\Om$. Moreover,  $\F$ must include the
set  $\Om$ itself and the empty set  $\emptyset$. Also, the set
$\F$ must be closed with respect to the complement operation, and
with respect to the countable unions and intersections of its
elements. A set is closed with respect to a particular operation,
if as a result of its application we obtain another element of the
set.

Finally, $P$  is a probability measure. This means, that a number
$P(F)$ is assigned to any event  $F\in\F$, and the following
conditions are satisfied: a)  $0\leq P(F) \leq 1$ for any

$F\in\F$, $P(\Om)=1$; b)  $P(\sum_j F_j)=\sum_j P(F_j)$ for any
countable family of non-overlapping subsets
 $F_j\in \F$.

Note that the probabilities  $P(F)$ are defined only for $F\in\F$.
If $F\notin\F$ then generally speaking the probability of such
subset  $F$ may not exist. This is also true for the elementary
events themselves.

From the physics point of view the choice of a particular
$\sss$-algebra $\F$ is determined by selection of measuring
devices for use in experimental investigations. These devices must
be able to distinguish one event in  $\F$ from another. In this
respect, quantum systems are qualitatively different from
classical ones. Because of the existence of incompatible
observables in the quantum case, measuring devices can distinguish
only those events, which differ from each other by the values of
observables  $\A$ from a single subalgebra  $\QQQ$.

Thus, a particular $\sss$-algebra $\F_{\xi}$ and a particular
system of probability measures  $\{P(F)\}_{\xi}$ where
$F\in\F_{\xi}$, corresponds to each subalgebra $\QQQ$. It is very
important that in the case of quantum systems there is no
$\sss$-algebra $\F_0$ having the following properties. A
probability measure $P(F)$  is defined for all events $F\in\F_0$,
and all algebras $\F_{\xi}$ are subalgebras of the algebra $\F_0$.
In other words, the probability distributions corresponding to
different  $\sss$-algebras $\F_{\xi}$ are not marginal
distributions of some general probability distribution. If we do
not take into account these special features of quantum systems,
then using the classical probability theory (see, e.g.,
\cc{chsh}), one can easily obtain for them the Bell inequalities
\cc{bel1,bel2}, which contradict the experimental data. On this
subject, see  \cc{slav1}.

Now we are going to show how the average value in a quantum state
$\Psi_{\vp_{\eta}}=\{\vp\}_{\vp_{\eta}}$ is constructed for an
observable $\A$. Let the physical system under consideration be in
the elementary state $\vp\in\{\vp\}_{\vp_{\eta}}$. If a device
$\xi$-type is used for measuring the value of an observable $\A$
($\A\in\QQQ$) then the results will be $A_{\xi}=\vx(\A)$
($\vx\in\vp=[\vx]$).

Let us now considered the event  $F_A$ containing all elementary
events $\vp$, such that  $\vp_{\xi}(\A)\le A$ ($\vp_{\xi}\in\vp$).
Denote $P(\vp: \vp_{\xi}(\A)\le A)$ the probability measure
corresponding to the event $F_A$ . We also introduce the notation

$$P_{\A}(d\vp)=P(\vp:\vp_{\xi}(\A)\leq A+dA)-P(\vp:\vp_{\xi}
(\A)\leq A)$$

 The physical meaning of  $P_{\A}(d\vp)$ is the probability
that a measurement of the observable $\A$ yields a value $A_{\xi}$
satisfying the condition $A<A_{\xi}\leq A+dA$. Then, the mean
value of the observable  $\A$ in the quantum state
$\Psi_{\vp_{\eta}}$ is given by the formula
  \beq{3}
\Psi_{\vp_{\eta}}(\A)=\int_{\vp\in\Psi_{\vp_{\eta}}}\,P_{\A}(d\vp)\,\vx(\A).
  \eeq

The average value of the observable  $\A$ defined above may depend
on the type of the measuring device used in the experiment (in the
case when the observable $\A$ belongs to several $\QQQ$
simultaneously). However, the experiment shows that such
dependence is actually absent. Therefore, we have to accept the
following postulate.

\

\post{} 6.  If  $\A\in\QQQ\cap\qqq$ and $\vp
\in\Psi_{\vp_{\eta}}$, then for  $\vx\in \vp$ and $\vp_{\xi'}\in
\vp$ we have
 $$ P(\vp:
\vx(\A)\le A)= P(\vp:\vp_{\xi'}(\A)\le A)$$.

\

Therefore, Eq. \rr{3} can be rewritten in the following form
 \beq{5}
\Psi_{\vp_{\eta}}(\A)=\int_{\vp\in\Psi_{\vp_{\eta}}}\,P_{\A}(d\vp)\,\vp(\A)
=\int_{\vp\in\Psi_{\vp_{\eta}}}\,P_{\A}(d\vp)\,A(\vp).
  \eeq
Here, firstly $A_{\xi}(\vp)\equiv\vx(\A)$, secondly, it is assumed
that one can substitute any $\vx$  instead of  $\vp$, and any
$A_{\xi}(\vp)$  instead of  $A(\vp)$.

In order for Eq. \rr{5} to correctly describe the mean values of
quantum observables one has to accept one more postulate.

\

\post{} 7. The probability distribution corresponding to a quantum
state is such that for any  $\A$ and $\B$ from $\AAA_+$ we have
 $$\Psi_{\vp_{\eta}}(\A+\B)=\Psi_{\vp_{\eta}}(\A)+\Psi_{\vp_{\eta}}(\B).$$

\ It is shown in the paper  \cc{slav1} that such distribution
actually exists.

The functional $\Psi_{\vp_{\eta}}(\A)$ defined by \rr{5} has a
unique linear extension on the entire algebra~$\AAA$:

$$\Psi_{\vp_{\eta}}(\D)=
\Psi_{\vp_{\eta}}(\A)+i\Psi_{\vp_{\eta}}(\B),
$$
where  $\D=\A+i\B$, $\A\in\AAA_+$, $\B\in\AAA_+$. Since $\vx(\A)$
is a character of a commutative associative algebra, the
functional  $\Psi_{\vp_{\eta}}$ is positive, and it is normalized
by the condition $\Psi_{\vp_{\eta}}(\I)=1$, where $\I$ is the unit
element of the algebra~$\AAA$.

In other words, the functional defined by  \rr{5} is a quantum
state, in the sense specified within the algebraic approach. Note
that the frequently used definition of a quantum state as a vector
in some Hilbert space, or a density matrix is particular cases in
the definition  of the algebraic approach.

Given a   $C^*$-algebra  and a linear positive normalized
functional defined there one can construct a Hilbert-space
representation of this algebra using the Gelfand-Naimark-Segal

(GNS) canonical construction (see, e.g., \cc{emch,naj}). Briefly,
the essence of the GNS construction can be described as follows.

Consider some  $C^*$-algebra \AAA{} and a functional  $\Psi$,
having the properties mentioned above. Two elements $\U$ and $\U'$
of the algebra  $\AAA$ are called equivalent, if for any
$\W\in\AAA$ we have $\Psi\left(\W^*(\U-\U')\right)=0$. Denote
$\Psi(\U)$, the equivalence class of the element  $\U$, and
$\AAA(\Psi)$ the set of all equivalence classes in  \AAA. The set
$\AAA(\Psi)$ is converted into a linear space if we define linear
operations there by the formula
$a\Phi(\U)+b\Phi(\V)=\Phi(a\U+b\V)$. Let us define now the scalar
product by the formula
  \beq{6}
\left(\Phi(\U),\Phi(\V)\right)=\Psi(\U^*\V).
  \eeq

The completion of the space  $\AAA(\Psi)$ with respect to the norm
$\|\Phi(\U)\|=[\Psi(\U^*\U)]^{1/2}$ turns  $\AAA(\Psi)$ â into a
Hilbert space  $\HHH$. Note that in contrast to $\HHH$, the set of
elementary states is not a linear space. In particular the notion
of "the sum of two elementary states" does not have a physical
meaning.

Each element  $\V$ of the algebra  \AAA{} is uniquely represented
in the space  $\HHH$ by a bounded linear operator $\Pi(\V)$,
defined by
  \beq{7}
\Pi(\V)\Phi(\U)=\Phi(\V\U).
  \eeq

According to Eqs. \rr{6} and  \rr{7}
 $$
 \lt(\Phi(\I),\Pi(\U)\Phi(\I)\rt)= \Psi(\U)
 $$

Thus, there are two paths leading to the same result. One can fix
the algebra of observables, and build on it a set of elementary

states corresponding to some quantum states. Then, one can endow
this set by the structure of a probability space and, finally,
calculate the probabilistic averages.

The alternative path is the following one. Fix a Hilbert space,
define observables as linear operators in that space, while
quantum states are either vectors of that space, or density
matrices. The average values of observables are defined as the
mathematical expectations of the corresponding operators with
respect to either vectors of the Hilbert space, or density
matrices.

Usually the second path turns out to be much more convenient from
the pragmatic point of view. However, the first path has a better
physical foundation and is much clearer. Therefore, it is better
suited for physical modeling. Moreover, one has to bear in mind
that the two specified paths are equivalent only when we consider
ensembles, which correspond to the classes of equivalent
elementary states described above or to linear mixtures of such
ensembles. Below we shall call them the quantum ensembles. This is
a very important type of ensembles, but it is far from being the
most general one. In particular, it does not contain ensembles
consisting from a single elementary state. Therefore, the
relevance of the standard mathematical formalism of quantum
mechanics to individual physical systems is rather questionable.
Later we will return to this problem.

On comparison of the two paths mentioned above it may seem that at
least in one particular aspect the second path has a wider domain
of applicability. The $C^*$--algebra used in the first path is a
Banach algebra, that is, all its elements have finite norms.
Therefore, this algebra describes only bounded observables. At the
same time, linear operators in a Hilbert space can be both bounded
and unbounded. This allows one to use them for description of not
only bounded observables, but also of unbounded ones. The latter
are widely used in quantum mechanics.

However, observables are described by self-adjoint operators, and
any such operator  ${\cal A}$ (bounded or unbounded) can be
uniquely represented in the form of its spectral expansion

  \beq{9}
{\cal A}=\int_{\sss({\cal A})}\lll\,P(d\lll).
  \eeq
Here, the numerical parameter  $\lll$ runs over some subset of the
real axis (the spectrum  $\sss({\cal A})$ of the operator  ${\cal
A}$), $P(d\lll)$ is a projector-valued measure on the spectrum
$\sss({\cal A})$. The latter means, that a projection operator
$P(d\lll)$ is assigned to each subset $d\lll$ of the spectrum
$\sss({\cal A})$. However, any projection operator has a finite
norm. Therefore, any operator describing an observable can be
expressed in terms of bounded operators.

The corresponding procedure can be applied in the framework of an
abstract $C^*$-algebra as well. In this case, the term "projector"
is used instead of the term "projection operator," while the
obtained element is called an adjoined element of the
$C^*$-algebra.

The measurement of an observable represented by Eq. \rr{9}
deserves a special discussion. Schematically, the corresponding
experiment looks like this. An appropriate measuring device is put
in contact with the system under investigations. As a result of
such contact the device "pointer" either points towards some
singled out interval on the device's scale, or it does not. Such
measurements are called "yes-no experiments." With the help of a
combination of such experiments one can determine the value of the
observable we are interested in with the required precision.

A special observable can be associated with each "yes-no
experiment." It obtains the value 1, if the pointer points towards
the interval singled out on the scale and the value 0 if it does
not. Such an observable has the properties of a projector $\p$,
that is, the properties of a Hermitian element of $C^*$-algebra,
which satisfies the condition $\p^2=\p$. Thus, a measurement of
any observable can be reduced to a combination of measurements of
observables described by projectors  $\p$. This procedure can be
considered as a physical realization of Eq. \rr{9}.

\section{LOCAL PROPERTIES OF OBSERVABLES AND STATES}

Now we will to try to apply the general discussion from Section 2
to solving the locality problem. By the locality we mean the
concentration in a domain of the four-dimensional Minkowski space,
uniting the time and the three spatial coordinates. We begin with
the discussion of local observables.

We will assume that an observable  $\A$ is localized in a domain
$\OO$ of Minkowski space if its value can be determined with the
help of measuring devices concentrated in this domain.

Let's ask ourselves the question, if a coordinate in the
nonrelativistic quantum mechanics is a local observable? At first
the answer seems trivial. Yes. However, this is not the case. In
that same nonrelativistic quantum mechanics it is claimed that the
time is not a local quantum observable. Therefore, in relativistic
quantum mechanics we have to assume that coordinates are also not
local quantum observables. However, there is no sharp boundary
between relativistic and nonrelativistic quantum mechanics. In
contrast, there is no any intermediate statement between the
assertions that the coordinates are local quantum observables, and
that they are not local quantum observables. Therefore, in order
to be consistent we cannot classify the coordinates as local
quantum observables in nonrelativistic quantum mechanics as well.

What can be used instead of the observables usually called the
coordinates? In order to answer this question we have to go back
to the discussion at the end of the previous section. Rulers are
commonly used as devices measuring the spatial characteristics of
objects. When measuring spatial coordinates of an object (a point
object, for simplicity) we actually state that the object is
located between certain points of a ruler. Thereby we state that
the value of the corresponding projector is equal to 1. It is this
projector that is the quantum observable, which characterizes the
spatial coordinate of the object. This observable is a local one.
Its domain of localization in the three-dimensional coordinate
space is bounded by the corresponding interval between neighboring
points on the ruler.

What is the role of the (classical) coordinate in the Minkowski
space? It distinguishes one interval on a ruler from another. In
other words, it distinguishes one quantum observable (a projector)
from another. Nevertheless, it may seem that a classical
coordinate still characterizes the localization of a quantum
state. Indeed, let us build an ensemble of physical objects (below
we shall call it a quantum ensemble) corresponding to the quantum
state we are interested in as follows. We will include in the
ensemble all the particles, which at the time of measurement were
found, for instance, in the interval number 3. It seems like in
doing that we have constructed a quantum state concentrated around
a certain classical coordinate.

However, in reality we have constructed only a sample from the
ensemble corresponding to that quantum state. Nothing prevents us
from using a second ruler, located in an entirely different domain
of the Minkowski space, together with the first one. The ensemble
constructed with the help of the first ruler can be complemented
with the particles, which happened to be in the interval number 3
on the second ruler. By its quantum-statistical characteristics
the new ensemble will look even better than the first one. It will
be a more representative sample corresponding to the same quantum
state. At the same time, the new sample will be localized now in
two domains of the Minkowski space. It is clear that this process
of increasing the number of utilized rulers can be continued as
long as necessary.

Thus, we see that the connection of a classical coordinate with
the quantum observable, which is usually called the coordinate, is
very indirect. Firstly, classical coordinates characterize the
localization of the utilized classical devices. Secondly, they
identify the observables of interest to us from a bigger class of
observables associated with every such measuring device, for
instance, with the singled out interval on each of the rulers used
in the experiment. Such a situation is explained by the fact that
construction of a quantum state is connected not with a particular
classical device, but with a particular measuring procedure. This
procedure can be accomplished by many classical devices.

At this point it is appropriate to go back to the Tonomura
experiment mentioned in the introduction. It seems that the
obtained interference pattern characterizes the spatial
distribution of scattered electrons. However, if we take into
account that during the time necessary for forming this picture
the measuring device together with the Earth traveled over a huge
distance, then in the coordinate system connected with static
stars one will fail to see any interference picture in the
distribution of electrons.

An analogous result can be demonstrated without a reference to
static stars. For this purpose it is sufficient to manufacture a
large number of copies of the Tonomura device and put them in
different places. Then, register a small number of scattered
electrons on each of these devices. We are not going to see the
interference pattern on any of the monitors of these devices.
However, if we collect the pictures from every monitor and combine
them, then, provided the number of devices is sufficiently large,
we will see a clear interference pattern. This demonstrates that a
quantum state associated with a certain interference picture
exists outside time and space.

Let us say now a few words about the spatial sizes of quantum
objects. One can frequently hear that quantum objects have sizes
of the same order of magnitude as the de Broglie wavelength
associated with this object. However, as was already mentioned in
the introduction, according to the point of view adopted here, the
de Broglie wavelength is a characteristic of an ensemble of
quantum objects, but not of a single object. Moreover, if we
connect the size of a quantum object with the de Broglie
wavelength, we arrive at paradoxical results of the following
type: the size of an electron at rest is infinite. Of course, one
can try to resolve this paradox with the help of vague reasoning,
that because of the uncertainty principle our ideas about sizes of
physical objects based on the classical intuition are no longer
applicable to quantum objects.

However, there is also another way to follow. Consider the
observables $\p_1$, $\p_2$, and $\p_3$ associated with three
consecutive intervals of a ruler. These observables are
compatible. Therefore they can be measured simultaneously. If such
a measurement yields  $p_1=0$, $p_2=1$, and $p_3=0$, then we can
conclude that the size of the object under investigation does not
exceed the length of the ruler's interval. In principle, such a
measuring procedure can be applied to any individual object. It
does not contradict our classical intuition.

Now we are going to discuss the locality properties of elementary
states. An elementary state is an attribute of an individual
physical system. The latter exists both in space and in time.
Therefore, one can expect that in contrast to quantum states, an
elementary state does possess the property of being localizable.

An elementary state is a mathematical model of material causes,
which determine one or another result of measuring the observables
of a physical system under investigation. Therefore, it is natural
to assume that an elementary state, restricted on the observables
which are defined in the domain $\OO$, is localized in this
domain. However, the actual situation is somewhat more
complicated.

In order to illustrate the problems, which arise here we consider
the process of elastic scattering of an electron on a nucleus.
This process is well studied both theoretically and
experimentally. Since an electron is much lighter than a nucleus,
this process is well approximated by the electron scattering on a
classical potential center (see, e.g.,  \cc{pesk}).

\begin{figure}[h]
 \begin{center}

  \includegraphics{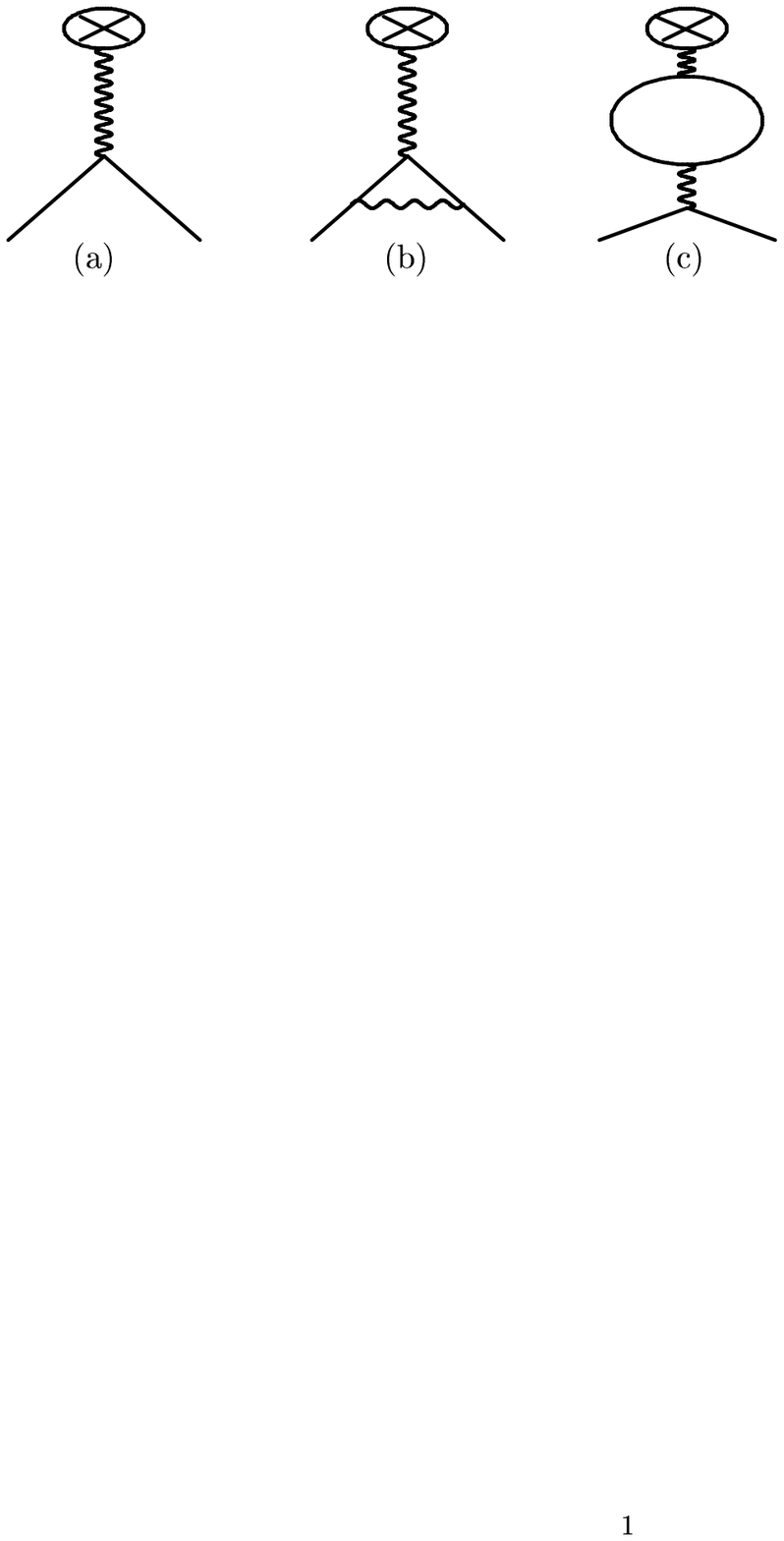}
 \includegraphics{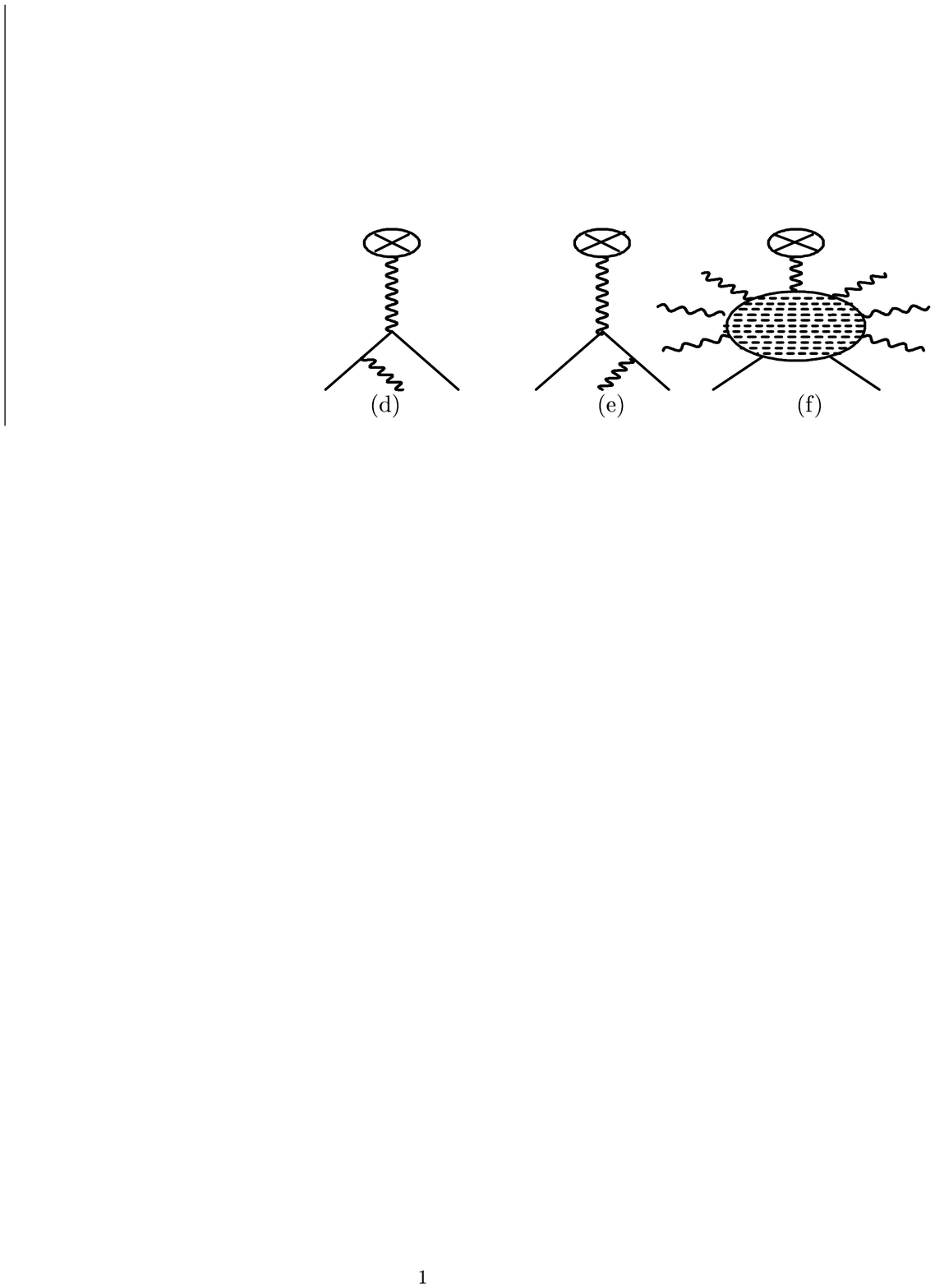}

\caption{ The Feynman diagrams describing the elastic scattering
of an electron (a-c) and the bremsstrahlung  (d-f). }
\end{center}
\end{figure}

In order to evaluate the cross section for elastic scattering of
an electron on a potential center in the lowest order of
perturbation theory we have to take into account the contributions
of the Feynman diagram  (a) shown in Fig. 2. The straight lines in
this picture correspond to the electron, the wavy lines
--- photons, and the circle with a cross --- the classical potential
center. The diagram  (a) does not lead to any difficulties. In
order to take into account the next order of the perturbation
theory we have to calculate the contributions from the diagrams
(b) and (c). Here, problems start to emerge. First of all, the
ultraviolet divergences appear. This problem can be solved with
the help of the standard renormalization procedure. Second, the
diagram  (c) leads to infrared divergences, which are typical for
processes involving massless particles. In this case --- photons.

In order to overcome this difficulty one has to take into account
that the elastic scattering process cannot be separated
experimentally   from   the   bremsstrahlung  process, where
bremsstrahlung photons with the energy below the sensitivity
threshold of the measuring device are being emitted. The Feynman
diagrams  (d), (e), and (f) in Fig. 2 correspond to this process.
In the experiment we always measure the total elastic scattering's
cross section and the cross section with bremsstrahlung photons
emission with energies less than certain value $E_{min}$. The
contributions to the total cross section from the diagrams (b),
(c), (d), and (e) correspond to the same order of perturbation
theory.

The infrared divergences compensate for each other in the total
cross section, while the cross section itself acquires the
dependence on  $E_{min}$, which in itself is justified from the
physical point of view. However, the dependence on  $E_{min}$
turns out to be singular, and the cross section tends to $-\infty$
as $E_{min}\to 0$. At the present time this phenomenon is deemed
to be an artifact connected with the application of perturbation
theory. Indeed, summation of higher orders of perturbation theory,
which take into account processes with emission of many
bremsstrahlung photons (diagrams of the type  (f)), transforms the
dependence on $E_{min}$ into a regular one, and as  $E_{min}\to 0$
the cross section of the process, which we interpret as elastic,
also tends to zero.

This example shows that depending on the characteristics of the
classical measuring device (in particular, depending on its
sensitivity) we can give different interpretations to the same
physical process. In the considered case it is either purely
elastic electron scattering, or electron scattering accompanied by
bremsstrahlung. At the same time we will have different ideas
about the post-scattering localization of the system.

If the measuring device allows one to register only electrons,
then it is natural to assume that the localization domain of the
physical system is the area where that electron is registered. If
for the same system under investigation the measuring device also
allows one to register a part of bremsstrahlung photons, then one
should include in the localization domain of the physical system
the area where these photons are registered. Thus, we have an
ambiguity in what we understand by the term, "the localization
domain of a physical system."

It is desirable to explicitly express this ambiguity in the
terminology, instead of "hiding it under the carpet." One can do
that by introducing the notions of the "kern" and "dark field" of
a physical system. The measuring device reacts on the part of the
investigated physical system called kern. Therefore the, kern
localization coincides with the localization of the registered
observables. The dark field is the part of the investigated
system, which the measuring device does not notice. Separation of
the investigated system on the kern and the dark field is not
unique. It depends on the characteristics of the measuring
equipment. However, in any case, a physical system has both the
kern and the accompanying dark field.

Let us go back to the concrete example considered above. There,
$E_{min}$ is the parameter characterizing the boundary between the
kern and the dark field. On the other hand, as was already
mentioned above, the elastic scattering's cross section of the
electron, that is, of that part of the physical system, which was
attributed to the kern, depends on  $E_{min}$. We also assumed
earlier that elementary states determine the results of
measurements, which can be performed on a physical system. Since
$E_{min}$ is determined by the total energy of all bremsstrahlung
photons (the dark field energy), we should include the domain of
the dark field's localization in the domain of the elementary
state's localization.

Note that relative to the scattering domain both the scattered
electron and the bremsstrahlung photons propagate in the future
light cone. Therefore, the inclusion of the dark-field
localization's domain in the localization domain of the electron's
elementary state does not contradict the locality conditions
accepted in the quantum field theory.

However, one can raise the following objection to the above
argument. Before the last scattering acts the electron was also
taking part in some other interactions. Each of those interactions
was also accompanied by emission of the bremsstrahlung photons,
which propagate in another light cone. It seems that those photons
should also be taken into account when we determine the
localization of the electron's elementary state. By continuing
this process of going deeper into history one can arrive at the
conclusion that the entire Universe should be included in the
localization domain of the electron's elementary state. One can
avoid such a paradoxical conclusion if we assume that every
interaction act is accompanied not only by creation of a new dark
field, but also by forgetting the previous history. One can
propose the following mechanism for realization of such a
scenario.

Let the measurement results depend only on the kern structure and
on the part of the dark field coherent to the kern. Then, the
elementary state will be determined only by these two structures.
Each interaction act is accompanied by a restructuring of the
system. As a rule, but not always, the newly formed structure
turns out to be incoherent with the previous structure.
Accordingly, the new elementary state turns out to be unrelated to
the previous structure.

Thus, it looks like the previous structure ceases to exist for the
new elementary state of the system under investigation. This is
how the forgetting process takes place. Of course, this does not
mean that the previous structure must disappear completely. The
part of the dark field, which is not involved in the interaction,
continues its previous existence. However, now that part of the
dark energy is no longer coherent to any kern. Therefore it cannot
show up in any quantum experiment. Such a dark field seems to be a
good candidate on the role of the dark matter's ingredient.

At the same time such an independently existing dark field can
show up as some classical field. For instance, in order to avoid
the appearance of infrared singularities in the process of
electron scattering on a potential center discussed above, one has
to take into account the possibility of emission of an infinite
number of photons having a finite total energy. Such a family of
photons will have all the properties of the classic
electromagnetic field. That opens the possibility for the
existence of a classical field, which is not an approximation of a
quantum field.

In the electromagnetic field case such classical field is a
classical tail of a massless quantum field. In principle, one can
assume the existence of a tail without a quantum analogue. The
gravitational field can pretend of the role of such an independent
classical tail.

Summarizing this section we would like to state the following. A
quantum has no localization. An elementary state has a finite
domain of localization. Somewhat arbitrarily an elementary state
can be divided into two parts: the kern and the dark field. The
arbitrariness of this separation comes from the fact that any
physical object has infinitely many characteristics, which can be
considered as observables. However, in any real measurement,
depending on the sensitivity of the measuring device, we consider
as observables only a certain part of these characteristics. It is
with these observables that the kern is associated. Thus, the
localization of the kern coincides with the localization of the
observables that we take into account. The observables not taken
into account become collectively a part of the dark field. This
dark field also has a finite localization domain, which is,
however, larger than that of the kern.

\section {DOUBLE-SLIT SCATTERING,\\
THE DELAYED CHOICE EXPERIMENT}

The notions of kern and dark field allow one to give a very
intuitive interpretation to the experiment, where a quantum
particle (for instance, an electron) is scattered on a double-slit
screen. According to Feynman \cc{feyn1}, this is "...a phenomenon
which is impossible, absolutely impossible, to explain in any
classical way, and which has in it the heart of quantum
mechanics."

Despite such a categorical statement by Feynman, we shall try to
give, if not an entirely classical, then such an explanation to
this phenomenon, which entirely fits the framework of classical
logic.

In a double-slit experiment on the electron beam's scattering a
clear interference pattern is observed. The Tonomura experiment
proves that this pattern is entirely determined by the interaction
of individual electrons with the slits, more specifically with the
screen where these slits were cut, or with the biprism, as it
actually happened in the Tonomura experiment. This experiment also
proves that the interference pattern is a purely statistical
effect. The interference pattern appears only when a sufficiently
large number of electrons are registered.

According to the classical ideas every individual electron passes
throw either one or another slit. In this case, if we are
alternately closing the slits one after another, we simply slow
down the accumulation of the statistical data, and we have to wait
longer before the interference pattern appears. However, this does
not happen, the interference pattern does not appear at all. This
means that when passing through one of the slits every electron
feels, whether the other slit is open or closed. On the other
hand, if every electron passes through only one slit, it is
natural to think that at the time instant of passing through the
screen it is localized in the domain of that slit. Then, in this
moment it belongs to the domain, which is space-like separated
with the second slit. Nevertheless the electron feels its state.
It would seem a violation of the locality postulate is taking
place here.

One can avoid such a violation by assuming that the physical
system scattered on the slits consists of a kern and a dark field.
It is the kern that in this case we interpret experimentally as an
electron. That is, the kern is localized in one of the slits. The
accompanying dark field is localized in a wider domain,
encompassing both slits. Arriving at the screen where the slits
were cut, the dark field excites collective oscillations there.
The oscillations are very weak; however, they are coherent to both
the dark field and the kern. Therefore, these oscillations can
interact resonantly with the kern. That is, they can play the role
of a random force, which forms the momentum distribution of the
scattered electrons. Since the oscillations are collective, their
structure depends on the screen structure, which in turn depends
on whether one or both slits are open.

Let us now turn our physical discussion into a mathematical form.
For simplicity we shall assume that a homogeneous electron beam
impinges perpendicularly on a screen with two identical slits  $a$
and $b$. It is clear, that the interference pattern is formed by
electrons, which pass either through slit $a$, or through slit
$b$. The interference pattern itself is determined by the
probability distribution of the momentum of the electrons
scattered on the slits. Thus, the following three events are
essential for the description of this process. The event $F_a$ ---
the fact of electron passing through the slit  $a$, the event
$F_b$
--- the fact of electron passing through the slit  $b$, and the event  $F_k$ ---
the fact that the momentum of the scattered electron belongs to a
fixed small solid angle  $d\om$ around the direction ${\bf k}$.

We face a physical problem, which can be formulated as a typical
problem of finding the conditional probability of the event  $F_k$
given that the event  $F_a$ has happened, or the event $F_b$.
Denote $P(F)$ as the probability of the event  $F$. According to
the standard formulas of the probability theory (see, e.g.,
\cc{nev}), we obtain

\beq{10}
 P(F_k)=\frac{P(F_k\cap(F_a+F_b))}{P(F_a+F_b)}.
 \eeq
Here,  $F_1\cap F_2$ denotes the simultaneous realization of the
events $F_1$ and $F_2$. Since the beam is homogeneous and the
slits are identical, then  $P(F_a+F_b)=
P(F_a)+P(F_b)=2P(F_a)=2P(F_b)$. Taking this into account, one can
rewrite Eq. \rr{10} as follows

\beq{11}
 P(F_k)=\frac{1}{2}\frac{P(F_k\cap F_a)}{P(F_a)}
 +\frac{1}{2}\frac{P(F_k\cap F_b)}{P(F_b)}.
  \eeq
The first terms in the right hand side of Eq. \rr{11} describe the
probability of an electron falling in the solid angle $d\om$ after
being scattered by the  $a$, the second term is the same
probability after being scattered by the slit $b$. This equation
corresponds to the absence of interference.

When deriving the above result we have made a typical mistake,
which is committed when standard formulas of the probability
theory are applied to description of quantum processes. When
writing down Eq. \rr{10} we implicitly assumed that the
probability measure $P\lt(F_k\cap(F_a+F_b)\rt)$ exists. However,
due to the incompatibility of the observables corresponding to the
momentum and the coordinate, the event  $F_k$ on the one hand, and
the events  $(F_a$ and $F_b)$, on the other, are not compatible.
Therefore, one cannot assign any probability to the simultaneous
realization of the above events. That is, we made a mistake in the
very beginning of our derivation. One has to point out, that such
a mistake is very widespread. It is made during the derivation of
the Bell inequality (see, e.g.,  \cc{chsh}), during the proof of
the Kochen-Specker "no-go" theorem \cc{ks}, and during the proof
that quantum mechanical predictions contradict the local realism
\cc{ghz}.

Thus, a straightforward application of Eq. \rr{10} for the
description of electron scattering on a double-slit plane is not
acceptable. At the same time one can propose a way around that
problem. One has to assume that the above process contains two
stages. The first one --- the arrival of an electron (electron's
kern more specifically) either in the slit $a$ or $b$  --- should
be interpreted as a preparation of a new quantum state. In the
second stage one should use this quantum state as a new set of
elementary events. In this case the event  $F_k$ can be already
considered as an unconditional one.

Let us associate the observable $\p_a$   with the event $F_a$.
This observable assumes the value  $p_a=1$, if the electron finds
itself in the slit  $a$, and the value  $p_a=0$, if it does not.
Such an observable has all the properties of a projection
operator. Analogously one can introduce the observable $\p_b$. A
double-slit screen selects only those elementary events, for which
the observable $\p_a+\p_b$ has the value 1. Denote by the symbol
$\Psi_{a+b}$ the equivalence class of such elementary states, and
the quantum state corresponding to this equivalence class.
Generally speaking this state can be a mixed one, however, even in
this case the functional $\Psi_{a+b}(\cdot)$ corresponding to this
quantum state must be linear. The following identities are valid
for $\vp\in\Psi_{a+b}$:
 \beq{12}
 \vx(\I)=1,\quad \vx(\p_a+\p_b)=1, \quad \mbox{for } \vx\in\vp.
 \eeq
Now we compute the mean values of the observables  $\I$ and
$\p_a+\p_b$ using Eq. \rr{3}.  Taking into account Eq. \rr{12}, we
obtain
 \beq{13}
 \Psi_{a+b}(\I)=\Psi_{a+b}(\p_a+\p_b)=1.
 \eeq
Since characters are positive functional, the functional
$\Psi_{a+b}(\cdot)$ is positive as well. Therefore, the following
Cauchy-Bunyakovski-Schwarz inequality is valid
 \beq{14}
\lt|\Psi_{a+b}\lt(\A(\I-\p_a-\p_b)\rt)\rt|^2\leq\Psi_{a+b}(\A^*\A)
\Psi_{a+b}(\I-\p_a-\p_b).
 \eeq
Due to Eq. \rr{13} the right hand side of Eq. \rr{14} is equal to
0. Therefore,
 \beq{15}
\Psi_{a+b}(\A)=\Psi_{a+b}\lt(\A(\p_a+\p_b)\rt).
 \eeq
Analogously,
  \beq{16}
\Psi_{a+b}(\A)=\Psi_{a+b}\lt((\p_a+\p_b)\A\rt).
 \eeq
Replacing $\A$ in \rr{15}   by $(\p_a+\p_b)\A$ and taking into
account Eq. \rr{16} one obtains
  \beq{17}
\Psi_{a+b}(\A)=\Psi_{a+b}\lt((\p_a+\p_b)\A(\p_a+\p_b)\rt).
 \eeq

Let us assign the observable $\K$ to the event  $F_k$  and use Eq.
\rr{17}. for this observable. Then, for the mean value of $\K$ one
obtains

  \beq{18}
\langle\K\rangle=\Psi_{a+b}(\K)=
\Psi_{a+b}(\p_a\K\p_a)+\Psi_{a+b}(\p_b\K\p_b)+
\Psi_{a+b}(\p_a\K\p_b+\p_b\K\p_a).
 \eeq
The first two terms in the right hand side of Eq. \rr{18} describe
the result of electron scattering separately on each of the slits
$a$ and  $b$, while the third term describes the interference.

Note that during the derivation of Eq.  \rr{18} we assumed that
the electron (the electron's kern) passed either through the slit
$a$, or through the slit  $b$, but we did not assume that it
passed through both slits simultaneously in some mysterious
fashion. At the same time, the functional $\Psi_{a+b}(\cdot)$,
which describes the scattered electrons, depends on the facts,
whether only one or both slits are open. As was already said
above, this can be explained by the fact that the electron
scattering is affected by collective oscillations, and their
structure depends on both slits.

Two facts are essential for the derivation of Eq.  \rr{18}. First,
the linearity of the functional $\Psi_{a+b}(\cdot)$, and second,
the incompatibility of the observable $\K$ with the observables
$\p_a$ and $\p_b$. If these observables are compatible, then the
third term in the right hand side of  \rr{18}, which is
responsible for the interference, vanishes.

The linearity of the functional  $\Psi_{a+b}(\cdot)$ was justified
by the fact that the slits play the role of classical devices
preparing a quantum ensemble. We have to interpret this fact as a
special boundary condition, replacing the microscopic description
of the electron interaction with the slits. We cannot present a
quantitative microscopic description of this process; therefore,
we restrict ourselves to a qualitative description of how the
electron passing through one slit can feel the presence of another
slit.

One has to point out that a microscopic description of the
electron interaction with the slits is also missing in the
traditional approach to quantum mechanics, and the presence of the
slits is taken into account by a boundary condition for the wave
function. Moreover, it is assumed that one cannot give an
intuitively clear qualitative explanation to this experiment.

It also remains unclear, why after a nonlocal interaction with the
slits the electron leaves only a point mark on the monitor. Of
course, a collapse of the electron's wave function is said to be
due to the interaction with the monitor. Nevertheless, a
convincing explanation of how this collapse can be reconciled with
the relativity theory is not presented. If, however, we assume
that the monitor registers only well-localized electron's kerns,
then the explanation becomes trivial.

Within the traditional approach the dual behavior of quantum
systems is usually explained by the influence of the environment.
For instance, it is often said that during interaction with the
slits the electron behaves as a wave, while during interaction
with the monitor it behaves as a particle. In order to examine
this line of argument, 30 years ago Wheeler proposed a Gedanken
Experiment  \cc{wheel}, which recently was performed in an almost
ideal form  \cc{jac}.

The layout of the experimental device is shown in Fig. 3. The
version proposed by Wheeler contained semitransparent mirrors
instead of the beamsplitter $BS_{in}$ and  $BS_{out}$. The
experimental device is a Mach-Zehnder interferometer with long
arms. In the real experiment they had the length of 48 m. On the
classical level the operating principle of the device looks very
simple.

\begin{figure}[h]
 \begin{center}

  \includegraphics{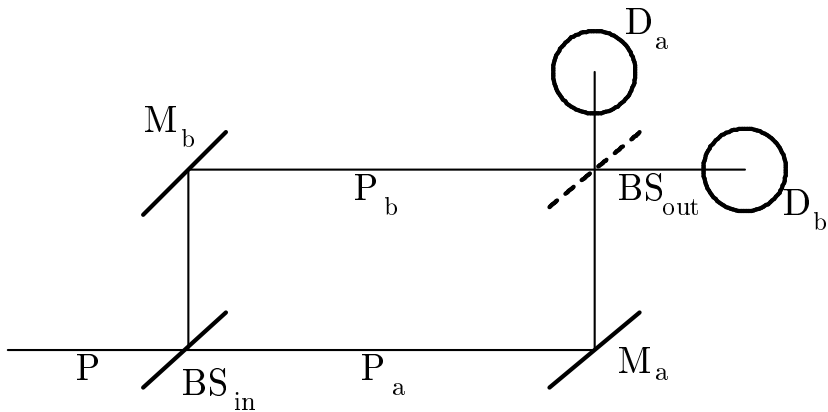}

   \caption{ The layout of the delayed-choice experiment. }
\end{center}
\end{figure}

A photon beam ($P$) falls on the entrance's semitransparent mirror
(beam splitter  $BS_{in}$). The mirror splits the beam into two
coherent parts  $P_a$ and  $P_b$, which follows the paths  $a$ and
$b$, and are reflected on their way by the mirrors $M_a$ and
$M_b$. There are two states of this device. The first one is when
the exit mirror (beam splitter  $BS_{out}$) is absent. In this
case, it is said the interferometer configuration is open, and
each part of the beam reaches the corresponding detector  ($D_{a}$
or $D_{b}$). In the second state the exit mirror is present (a
closed interferometer configuration). Both parts of the beam are
combined coherently by the exit mirror in this case. The result of
this combination depends on the fact that reflection of the beam
by the mirror changes its phase by $\pi/2$, while passing through
the mirror does not change the phase. Taking this into account it
is easy to see that after being combined in the exit mirror the
entire beam reaches the detector $D_b$.

The picture turns out to be much more interesting on the quantum
level. In order to obtain this picture in the pure form the beam
intensity is reduced sharply, so that not more than one photon can
be present in the device at any time. Each of these photons can
possess both corpuscular and wave properties. Let us assume that
depending on the state of the environment (the state of the
experimental device) photons exhibit either corpuscular or wave
properties. If a photon behaves as a corpuscular, then as a result
of interaction with the entrance mirror this photon randomly
selects one of the paths. If a photon behaves as a wave, then it
splits in the entrance mirror into two parts and follows the both
paths.

Let the state of the device be such that the exit mirror is
absent. Then, due to the corpuscular behavior of the photon, only
one of the detectors registers a signal. By registering which of
the two detectors has been triggered we can establish which of the
two paths has been selected by the photon in the entrance mirror.
If photons behave as waves, then the signals should be
simultaneously registered in both detectors. Now, let the state of
the device be such that the exit mirror is present. Then, if
photons behave as corpuscles just one of the detectors will be
triggered randomly as above. If they behave as waves then the
detector $D_b$ will be always triggered.

Thus, in order for the quantum picture to comply with the
classical one, the photon must behave as a corpuscle in the
absence of the exit mirror, that is, it must choose one of the two
paths. If however the exit mirror is present, then the photon must
behave as a wave, and it has to propagate over the both paths
after the entrance mirror.

It seems that the photon must choose either a single path or both
of them at the time of passing through the entrance mirror. In
order to "confuse" the photon, Wheeler proposed to take the
decision whether to put the exit mirror or not after the photon
passed through the entrance mirror, and to execute this decision
before the photon reaches the domain of the exit mirror. Thus, at
the time of passing through the entrance mirror "the actual
situation will not be clear" to the photon. Nevertheless, in order
to reproduce the classical picture, the photon must make the
correct choice every time, that is, it should be able to guess
beforehand the experimentalist's fads.

The experimental implementation of the mirror manipulation
proposed by Wheeler turned out to be very difficult. It was
necessary to get through everything in 160 ns --- the time taken
by the photon to pass through the interferometer base (48 m). The
experimentalists managed to complete all the required
manipulations in 40 ns. Of course, one cannot do that with a
semitransparent mirror. Therefore, a beam splitter was used
instead of the mirror, which was switched on and off with an
electro-optical modulator. The decision whether to switch the beam
splitter on or off was taken by a random number generator. The
device geometry was such that no signal traveling with a velocity
not exceeding the speed of light could pass the information about
the taken decision to the entrance mirror before the investigated
photon had reached it.

Despite all these precautionary measures the photon was able to
perfectly foresee the decisions taken by the random number
generator. This means that during the time interval between the
instances of passing through the entrance and exit mirrors the
photon does not have a choice whether to be localized in one of
the interferometer arms or in both of them simultaneously. In some
mysterious way both of these possibilities are realized
simultaneously. Formally, this does not contradict the standard
mathematical formalism of quantum mechanics. However, standard
quantum mechanics does not allow one to compose any intuitive
physical picture of this phenomenon.

In contrast, in terms of the elementary states, the kern and dark
field, the physical picture of this phenomenon looks very simple.
Reaching the entrance mirror the photon interacts with it.
Depending on the photon's elementary state its kern is either
reflected by the mirror, or goes through it. Simultaneously, a
dark field coherent to the kern is being created as a result of
the interaction. This dark field is separated into two parts, one
of which propagates through one path, another through the other
path. Thus, during the time interval mentioned in the previous
paragraph the electron's kern and one of the parts of the dark
field are localized in one shoulder of the interferometer, while
the second part of the dark field in the other shoulder. All parts
of the photon retain the coherence with each other.

In the exit mirror both parts of the dark field are combined
coherently, generating small secondary oscillations of the mirror
coherent to the kern. These secondary oscillations interact
resonantly with the kern. Due to the phase shift caused by the
interaction of the photon's components with the mirrors the
resulting dark field and the kern propagate toward the detector
$D_b$ after passing through the exit mirror. Once the kern hits
the detector the latter registers the arrival of the photon.

In the absence of the exit mirror the kern continues to propagate
along the path chosen at the entrance mirror and hits one of the
detectors triggering the photon registration. The part of the dark
field propagating over the other path reaches the other detector.
However, the detector does not respond to the dark field. In this
case, the entire picture looks as if the photon possesses only
corpuscular properties. Thus, the general picture looks quite
clear and agrees completely with the locality and the causality
principles.

\section{ENTANGLED STATES AND THE EPR PARADOX}

Upon discussing the locality problem the most interesting and
mysterious seems to be the topics related to the entangled states.
This term was once introduced by Schroedinger  \cc{schr}, and
originally it was "Verschrankung." For a two-particle system where
each of the particles can be in two orthogonal quantum states
$|+\ra$ and $|-\ra$ the examples of typical entangled states are
  \bea{19}
|\Psi^{(-)}\ra_{12}&=&\frac{1}{\sqrt{2}}\lt[|+\ra_1|-\ra_2-
|-\ra_1|+\ra_2\rt],\nn
|\Psi^{(+)}\ra_{12}&=&\frac{1}{\sqrt{2}}\lt[|+\ra_1|-\ra_2+
|-\ra_1|+\ra_2\rt],\nn
|\Phi^{(-)}\ra_{12}&=&\frac{1}{\sqrt{2}}\lt[|+\ra_1|+\ra_2-
|-\ra_1|-\ra_2\rt],\nn
|\Phi^{(+)}\ra_{12}&=&\frac{1}{\sqrt{2}}\lt[|+\ra_1|+\ra_2+
|-\ra_1|-\ra_2\rt].
 \eea
Here,  $|\cdot\ra_{12}$ denote the state vectors from the Hilbert
space of the two-particle system, while $|\cdot\ra_{1}$ and
$|\cdot\ra_{2}$ are state vectors of the first and the second
particle in the Hilbert space of the one-particle system,
$|\cdot\ra_{1}|\cdot\ra_{2}$ is the direct product of the
corresponding vectors. The quantum states shown in Eq. \rr{19} are
often called the Bell states.

The distinctive feature of the entangled state is that using
classical measuring devices one can prepare the corresponding pure
states of the many-particle (two-particle in the case of
Eq.~\rr{19}) system. However, even after that one cannot say what
is the pure quantum state of each particle involved in the system.
On the other hand, if subsequently one performs a measurement with
one particle only, then one can establish not only the pure state
of this particle, but also the pure quantum state of its partner,
which was not involved in the measurement.

For instance, if we know that a two-particle system occupies the
quantum state  $|\Psi^{(-)}\ra_{12}$, then one cannot say with
certainty in which of the two possible states  $|+\ra$ or $|-\ra$,
each particle is. However, if as a result of a subsequent
measurement with the first particle it would be established that
it occupies the state  $|+\ra_{1}$, then one can predict that the
measurement with the second particle would show with a probability
of 1 that it occupies the state $|-\ra_{2}$.

Such a state of affairs is nailed down within the standard
approach to the quantum mechanics by the so-called projection
principle  \cc{von}. According to this principle the measurement
performed over a part of the investigated physical system leads to
a change (reduction) of the quantum state of the entire system.
Not only the characteristic of the part of the entire system,
which was subjected to the action of the measuring device, can
change, but of the other part, which was not subjected to such an
action as well. For instance, in the example considered above, as
a result of the measurement of characteristics of the first
particle the state  $|\Psi^{(-)}\ra_{12}$ reduces (collapses) to
the state  $|+\ra_{1}|-\ra_{2}$.

As a recipe for the mathematical description of the measuring
device's influence on the quantum object, the projection principle
as a rule works very well. However, within the standard approach
one cannot give to this principle any intuitive physical
interpretation, which would agree with the relativity theory.

In his famous book \cc{von} von Neumann introduces the notion of
two types of action on a physical system. As a result of an action
attributed by von Neumann to the second type the quantum state
evolves according to the Schroedinger equation. This change obeys
the causality principle and it is unambiguously predictable. This
is how a quantum state of a system evolves when it interacts with
another quantum system or with a classical external field.

The action of a measuring device on a physical system was
classified by von Neumann as the first type of action. Such an
action changes the quantum state of a system in a random fashion
and, according to von Neumann, it is acausal. This looks very
strange, because any measuring device can be considered either
like some quantum system, or as an external classical field. The
only distinctive feature of the interaction between a measuring
device and a physical system under investigation is that as a
result of this interaction we obtain some information about the
system. In this connection von Neumann introduced the notion of
psychophysical parallelism. According to this principle the
experimentalist's inner man plays a crucial role in the
description of the first type action. Using this principle von
Neumann tried to explain the unusual properties of this type of
action.

In contrast to other arguments by von Neumann this one does not
seem to be convincing at all. Later there were many attempts to
justify the projection principle; however, all of them, to put it
mildly, seem to be rather questionable from the physical point of
view.

Below we are not going to discuss general features of the
projection principle (it can be found in  \cc{slav2}), instead we
will concentrate our attention on the problems arising when this
principle is applied in a concrete case, namely, in the case of
the so-called Einstein-Podolsky-Rosen (EPR) paradox.

In the original version  \cc{epr} the authors considered
paradoxical phenomena arising during measurements of coordinates
and momentums in a system of two correlated particles. Later
Bohm~\cc{bohm} proposed another physical model for demonstration
of the same paradoxical phenomena. From the standpoint of
principle the model proposed by Bohm is quite identical to the
model considered in the paper  \cc{epr}. At the same time the Bohm
model is much more convenient for discussions. Therefore, we will
consider the version of the EPR paradox proposed by Bohm.

Bohm proposed to consider the singlet state of a system consisting
from two particles with spin 1/2. The total spin of the system in
this state is equal to zero. Such a state can appear as a result
of positronium decay into an electron and positron. In the
Hilbert-space notations this state is described by the vector
$|\Psi^{(-)}\ra_{12}$ (see Eq. \rr{19}). In this case, $|+\ra_1$
denotes the quantum state of the first particle with the spin
projection onto a singled out axis  (axis $z$) equal to  +1/2, and
$|-\ra_1$ is the quantum state with the projection equal to
$-1/2$. Analogous notations are used for the states of the second
particle.

In the nonrelativistic case the spin projection on a singled out
axis is a good quantum number, and it is conserved during free
motion. Denote the observable corresponding to the spin of the
first particle by  ${\bf \s}_1$. For the observables corresponding
to the spin projections on the  $z$  axis and on the vector ${\bf
n}$ we will use the notations $\s_{1z}$ and  $\s_{1n}$. Analogous
notations will be used for the second particle. For the
corresponding total quantities we will use the same symbols but
without the subscripts 1 or 2. As in the general case, the values
of observables will be denoted by the same symbols as the
observables themselves, but without the "hat."

So, let us consider a positronium in a singlet state, which
disintegrates into two particles with spin 1/2. Let us wait until
the particles fly apart and are separated by a large distance, and
then measure the spin projection on the $z$ axis for the first
particle. According to the standard rules of quantum mechanics
these particles do not have a definite value of  $S_{1z}$ and
$S_{2z}$ before the measurement. An experiment can give us the
values  +1/2 or  $-1/2$ with equal probabilities. Let us assume
that we have obtained the value  $S_{1z}=+1/2$. The first particle
is said to acquire a definite value (+1/2 in this case) of the
spin projection on the $z$ axis as a result of the measurement.
Since the first particle was in contact with the measuring device,
this statement looks quite plausible.

However, if after the first measurement we measure the spin
projection on the same axis for the second particle, we will
obtain  $S_{2z}=-1/2$. Explanation of this fact turns out to be a
much more difficult problem. We will assume that the second
measurement was performed so quickly after the first one that no
signal was able to reach the place of the second measurement from
the place of the first one. If we believe in the locality
principle, then we must conclude that the first measurement could
not influence the result of the second one. Nevertheless, a strict
correlation is found between the results of the two measurements.

The existence of such correlation can be explained by the fact
that during the positronium disintegration the electron and
positron obtained definite values of the spin projection on the
$z$ axis. The exact values are not known, but we do know that
their sum is equal to zero. This explanation gives a reasonably
clear intuitive interpretation of the obtained results. However,
it contradicts the general concepts accepted in the standard
quantum mechanics. The point is that instead of measuring the spin
projection on the $z$ axis we could have measured the spin
projection on the $x$ axis. The result would be analogous: the sum
of the spin projections on the  $x$ axis in both particles is
equal to zero.

Therefore, in order for the above explanation to remain in force
we must assume that during the positronium disintegration both the
electron and the positron obtained definite values of the spin
projection not only on the  $z$ axis, but on the  $x$ axis as
well. However, for each particle the spin projections on the  $z$
and  $x$ axis are incompatible observables. Therefore, according
to the concepts accepted in the standard quantum mechanics they
cannot have definite values simultaneously.

This contradiction is the essence of the EPR paradox. Those who
deny the presence of the EPR paradox claim the following. The EPR
paradox appears because we are trying to find an intuitive picture
of how correlations between different parts of a complex system
appear. However, our intuitive notions are based on the experience
obtained in experiments with macroscopic objects, while in the
world of microscopic objects these notions may turn out to be
inadequate.

Formally, such a point of view is admissible. We may think that
there are complicated correlations between remote parts of a
complex system, and the origin of these correlations is beyond our
comprehension. It is even possible that they do not have any
origin at all (the absence of causality in the microscopic world).
We can only state that such correlations exist and they are
described by the corresponding quantum state.

At a certain stage of science development such a point of view may
turn out to be very productive. We notice certain regularities in
nature; we do not spend time and efforts on attempts to explain
them, but begin to exploit them intensively. However, sooner or
later a necessity to establish those more fundamental rules
arises. It seems that for the quantum theory this time has come.

The acknowledgement that unexplainable correlations between
distant parts of a physical system exist leaves in doubt the
locality principle accepted in the quantum theory. In turn, this
raises hopes for the possibility of new ways of information
transmission unrestricted by the stringent requirements of the
relativity theory. At the same time references are often made on
supposedly the proven nonlocality of quantum measurements. More
specifically, the references on the absence of local reality
determining the results of local measurements.

The concept of the elementary state adopted in this paper allows
one to give an obvious interpretation to the EPR paradox while
sticking with the validity of the locality principle for quantum
measurements. According to this concept any quantum state,
including the singlet state  $\Psi_{12}^{(-)}$, describes an
equivalence class of elementary states. In other words, it
describes not an isolated physical system, but an ensemble of such
systems. This ensemble is not localized in the Minkowski space.
Therefore, physically, one cannot demand the validity of the
locality property for the singlet quantum state.

The determining feature of the singlet state is the fact that a
measurement of the total spin's projection on any vector  ${\bf
n}$ yields $S_n=0$. Therefore, if we perform a measurement of the
observables  $\s_{1n}$ and  $\s_{2n}$ in a compatible way, then we
must always obtain
  \beq{20}
S_{1n}+S_{2n}=0.
 \eeq

The validity of this equation does not depend on the localization
domain of the observables   $\s_{1n}$ and $\s_{2n}$. It is only
important that these measurements are compatible. Assuming the
validity of the locality principle the latter will be certainly
true if we measure these observable in space-like separated
domains.

Since Eq. \rr{20} is valid for each individual measurement, the
singlet state is an equivalence class of elementary states with
Eq. \rr{20} being valid for each of them. This is not in conflict
with the situation where each of the particles within the
considered two-particle system is in a certain individual
elementary state. This state is a mathematical model of a local
reality, which determines the results of local measurements for
this particle. The elementary states of individual particles
making up a singlet pair do not have to form an equivalence class
corresponding to a certain pure quantum state of a single-particle
system.

The paper  \cc{slav1} describes all elementary states of a
particle with spin 1/2. Demonstrably one can view such a state as
a sphere with unit radius, where each dot is either black or
white, and all antipodal dots have different colors. Each such ESS
(Elementary State Sphere) describes the results of spin projection
measurements as follows. In order to find out, which value will be
obtained for the spin projection on the vector ${\bf n}$ of a
particle in some elementary state, we have to draw in the
corresponding ESS, the radius in the direction of the vector ${\bf
n}$. If the radius hits a black point, then the projection value
is +1/2, if it hits a white point, then the value is  $-1/2$.
Accordingly, the singlet state of a two-particle system can be
viewed as a pair of ESSs, which are negative copies of one
another.

In each individual measurement a classical device can measure the
spin projection on any (but only on one) direction  ${\bf n}$. An
interaction with the measuring device changes the painting of ESS
uncontrollably. Therefore, we can never obtain the complete
information on the painting of ESS with the help of these
measurements. That is, one cannot obtain complete information on
the elementary state of the particle.

At the same time if we know that a two-particle system is in the
singlet state, then, having measured the spin projection on the
vector  ${\bf n}$ for a single particle, we can obtain the
information about the spin projection on the same vector for the
second particle. This happens because the ESS's of these particles
are perfectly anti-correlated, as prescribed by Eq. \rr{20}. Such
measurement of the spin projection for the second particle is
called indirect. It is possible for any localization of the second
particle. Note that the elementary state of the second particle is
not changed as a result of the measurement. Although the indirect
measurement gives us certain information on the elementary state
of the remote object, this does not violate the interaction's
locality principle by any means.

Since the singlet state  $\Psi^{(-)}_{12}$ is usually used in
discussions of the EPR paradox, this quantum state is often called
the EPR-state in the modern literature, while the corresponding
two-particle system is called the EPR-pair.

The electron-positron pair is very convenient for theoretical
discussions of the EPR paradox. However, experimental realization
of such a physical system is very difficult. It is much easier to
realize such a singlet pair by optical means. In this case the
EPR-pair consists of two photons polarized in two mutually
orthogonal directions: horizontal (the state vector
$|H\ra\equiv|+\ra$) and vertical (the state vector
$|V\ra\equiv|-\ra$). The process where optical EPR-pairs are
obtained is called the type II parametric down conversion (see,
e.g., \cc{grw}). In the Russian-language literature the term
"spontaneous parametric scattering" (SPS) is more frequently used.

The main element of the SPS source is a nonlinear crystal, which
is irradiated by an ultraviolet laser. Due to collective processes
the laser photons are scattered in the crystal. As a rule, a
single incident photon generates one outgoing photon. However,
sometimes, with much smaller probability, scattering of a single
photon leads to emission of two photons. In this case the energy
of the incident photon is split almost equally between the two
emitted photons. One can create such conditions that these two
photons form an EPR-pair.

It is the application of the SPS method that allowed us to
actually realize the experiment  \cc{adr} demonstrating the EPR
paradox. Since the results of this experiment are in complete
agreement with the predictions of standard quantum mechanics, it
is deemed that these results support the absence of local reality
determining the behavior of localized parts of quantum systems. As
was shown above, this conclusion is absolutely unnecessary. The
existence of a local reality described by an elementary state of
the corresponding part of the quantum system also leads to
conclusions, which are in an agreement with the results of the
performed experiment.

\section{QUANTUM TELEPORTATION}

The ideas of a nonlocal character of quantum measurements raised
great expectations for the possibility of a principally new way of
information transmission (see, e.g., \cc{pqi}). In scientific
literature this technique was called "quantum teleportation."
Numerous experiments were conducted already, and their results
seem to support these hopes. The element of mystery surrounding
the notion of "teleportation" is still present in scientific
literature too. Below we will try to remove this veil of mystery,
see also \cc{slav4}.

One can understand the essence of the teleportation phenomenon by
considering the Gedanken Experiment, whose scheme is shown in
Fig.~4.

\begin{figure}[h]
 \begin{center}

  \includegraphics{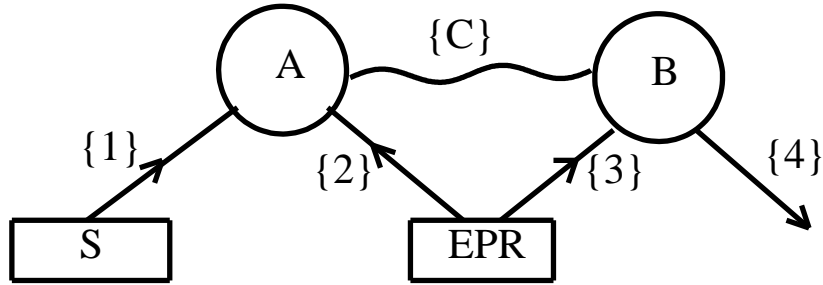}

   \caption{ The layout of quantum teleportation.}
\end{center}
\end{figure}

Here,  $S$ is the source of the initial state;  $EPR$ is the
source of EPR; $A$ is the Bell state's analyzer (Alice);  $B$ is
the unitary transformer (Bob); $\{C\}$ is the classical
communication channel; \{1\} is the carrier of the initial
teleported state; \{2\} -- \{3\} is the EPR pair; and  \{4\} is
the carrier of the final teleported state.

In the standard form the description of the teleportation
phenomenon looks as follows (see, e.g.,~\cc{bwz}).  The source $S$
emits a particle  \{1\} in the quantum state  $\Psi_1$ that is
described by the Hilbert space's vector
$|\Psi\ra_1=\al|+\ra+\beta|-\ra$, where $\al$ and  $\beta$ are
complex numbers satisfying the condition $|\al|^2+|\beta|^2=1$.
The particle  \{1\} is forwarded to Alice. The EPR source emits
the EPR pair  \{2\} and  \{3\} in the singlet state
$\Psi^{(-)}_{23}$ (the vector $|\Psi^{(-)}\ra_{23}$, see
Eq.~\rr{19}). One of the particles  (\{2\}) of the pair is
forwarded to Alice; the other  (\{3\}) is forwarded to Bob.

According to the standard rules of quantum mechanics the state of
the three-particle system (the particles  \{1\},\{2\}, and \{3\})
is described by the quantum state vector
$|\Psi\ra_{123}=|\Psi\ra_1|\Psi^{(-)}\ra_{23}$. This vector can be
represented as an expansion over the Bell states for the particles
\{1\} and  \{2\} (see Eq.~\rr{19}).
 \bea{21}
|\Psi\ra_{123}&=&\frac 12\lt\{|\Psi^{(-)}\ra_{12}\Big(-\al|+\ra_3
- \beta|-\ra_3\Big)+|\Psi^{(+)}\ra_{12}\Big(-\al|+\ra_3 +
\beta|-\ra_3\Big)\rt.\nn \
&+&\lt.|\Phi^{(-)}\ra_{12}\Big(\al|-\ra_3 +
\beta|+\ra_3\Big)+|\Phi^{(+)}\ra_{12}\Big(\al|-\ra_3 -
\beta|+\ra_3\Big)\rt\}.
 \eea
Using the analyzer  $A$ Alice determines, which of the four
possible Bell states describes the particle  \{1\} and \{2\}
forwarded to her. Lets say for instance that this is the state
$|\Psi^{(-)}\ra_{12}$. According to the projection principle of
standard quantum mechanics, after such a measurement the
three-particle state is reduced as follows
 \bea{22}
 |\Psi\ra_{123} \to|\Psi^{(-)}\ra_{12}\Big(-\al|+\ra_3 -
\beta|-\ra_3\Big).
 \eea
Using the classical communication channel Alice reports the
observed result to Bob. Having received the message that Alice
registered the state  $|\Psi^{(-)}\ra_{12}$ Bob does not do
anything, and lets the particle \{3\} go further. According to the
right hand side of Eq. \rr{22} this particle will be in the state
$|\Psi\ra_4=\Big(-\al|+\ra_3 - \beta|-\ra_3\Big)$.

The quantum states described by the vectors  $|\Psi\ra_1$ and
$|\Psi\ra_4$ coincide. Initially there were no correlations
between the quantum states of the particles  \{1\} and  \{3\}.
Alice interacted only with the particles \{1\} and  \{2\}. At that
time the particle  \{3\} could have been in a domain of Minkowski
space, which was space-like separated with the domain where
Alice's manipulations were taking place. Nevertheless, in some
mysterious way the state of the particle \{3\} became identical to
the quantum state of the particle \{1\}. Neither Alice, nor Bob
could know what the quantum state of the particle  \{1\} was.

If Alice finds that the particles  \{1\} and  \{2\} are in another
state, and reports her findings to Bob over the classical
communication channel, then Bob will have to do something. If the
observed state is $|\Psi^{(+)}\ra_{12}$, then Bob has to perform
an operation, which corresponds to the unitary transformation
$-|+\ra\longrightarrow|+\ra$. If the observed state is
$|\Phi^{(-)}\ra_{12}$, then the transformation $
|+\ra\longleftrightarrow|-\ra$ would be required. Finally, in the
case of the state  $|\Phi^{(+)}\ra_{12}$ the required
transformation $-|+\ra\longrightarrow|+\ra\longleftrightarrow
|-\ra$ would be required. After these operations the particle
\{3\} will be in the state $|\Psi\ra_4$ coinciding with the
quantum state $|\Psi\ra_1$. In the cases when Bob has to perform
some additional operations, the paradoxically of the situation
appears to be clouded. However, from the common point of view the
situation looks as absurd as in the first case.

The standard words which are pronounced in this situation sound
approximately as follows: "Many things, which seem absurd from the
ordinary point of view are quite normal in the quantum world."
However, one can do without such a poor consolation, if the
results of the experiment are interpreted using the notion of the
elementary state.

In the framework of this interpretation the fact that the source
$S$ emits particles in a certain quantum state means the
following. The source emits a beam of particles in different
elementary states; however, they all belong to the same
equivalence class. The beam does not have to be localized neither
in time, nor in space. It may not be known to anyone, to which
equivalence class those elementary states belong. This means that
the numbers  $\al$ and  $\beta$ in the expansion $|\Psi\ra_{1}$,
over the basis  $|+\ra$, $|-\ra$ may not be known, however, they
are the same for all the particles of the beam (up to a common
phase multiplier). Therefore, there exists such a coordinate
system in the three dimensional space where all the particles in
the beam will have the spin projection on the  $z$ axis equal to
+1/2. In this coordinate system, the value $\al|+\ra+\beta|-\ra$
corresponds to the state vector $S_z=+1/2$, $S_z=-1/2$ corresponds
to the vector $-\al|+\ra+\beta|-\ra$. One can find such a
direction for the $x$ axis, that $S_x=+1/2$ corresponds to the
vector $\al|-\ra+\beta|+\ra$, and $S_x=-1/2$ corresponds to the
vector $\al|-\ra-\beta|+\ra$.

Each particle in the beam emitted by the source  $S$, is analyzed
by Alice together with the particle of the EPR pair emitted by the
source  $EPR$. . Different EPR pairs are in different elementary
states, but in each pair the elementary state of one particle is
the negative copy of the elementary state of another particle. The
physical system consisting of the analyzer and the particle \{1\}
can be considered as a complex measuring device. Using this device
Alice sorts particles \{2\} over four groups. Each group
corresponds to one of the four quantum Bell states, and it unifies
particles  \{2\}, which together with the corresponding particle
\{1\} are in this two-particle elementary state.

Since each particle  \{2\} has a partner in the EPR pair, this
sorting can be considered as splitting the beam of  \{3\}
particles into four sub-beams. According to Eq. \rr{21} each of
these sub-beams will contain particles  \{3\} with a definite
value of the spin projection either on the  $z$ axis or  $x$ axis.
This happens because the elementary state of each particle \{3\}
is strictly correlated with the elementary state of the partner
particle \{2\}.

Since Alice determines to which group each of the particles \{2\}
goes, she also knows which group the corresponding particle \{3\}
falls into. Alice transmits this information to Bob via a
classical channel, which allows him to select the required unitary
transformation. With the help of this transformation he changes
the elementary state of the particle \{3\} in such a way that it
got through to the quantum state $\Psi_1$.

The measurements performed by Alice has no influence whatsoever on
the elementary state of the particle  \{3\}. She only receives
some information about that elementary state with the help of an
indirect measurement and shares it with Bob. After that Bob
performs certain manipulations with the particle  \{3\}. As a
result of these manipulations the elementary state of the particle
\{3\} does not become the exact copy of the elementary state of
the particle  \{1\}. Therefore, the term "teleportation" does not
seem to be quite appropriate in this case. Bob only managed to
drive the particle \{3\} in the same equivalence class where the
particle \{1\} was.

Next, we will discuss a real experiment where the quantum
teleportation was observed. This experiment was conducted with
photons, accordingly, optical devices where used. Before
proceeding with the discussion of the experiment we shall briefly
describe the operating principles of two devices utilized there.

The first optical device is a polarization beam splitter PBS. It
can split a photon beam into two sub-beams polarized in two
mutually orthogonal directions. The device's geometry determines
three orthogonal directions (an orthogonal polarization basis):
${\bf I}$ is the direction of the incident beam, ${\bf H}$ is the
horizontal direction,  ${\bf V}$ is the vertical direction. If the
incident photon beam is polarized horizontally, then after passing
though the PBS the photons propagate in the  ${\bf H}$ direction;
if the beam has vertical polarization, the photons propagate along
the  ${\bf V}$ direction. If the incident beam is polarized at a
certain angle  $\vt$ ($\vt\neq \pi n/2$), then a part of the
photons passing through the PBS will propagate in the  ${\bf H}$

direction and will acquire the horizontal polarization, the
remaining photons will propagate in the ${\bf V}$ direction and
acquire the vertical polarization.

In the context of a single photon in the beam this physical
phenomenon has essentially different interpretations within the
standard quantum-mechanical approach, and within the approach
described in Section 2.

Within the standard approach it is preferable not to talk about
the single photon's polarization at all. If it is known that a
photon belongs to a polarized beam, then it is said that the
photon has the corresponding polarization. If nothing is known
about the photon's previous history, then it is said that the
photon does not have any polarization. It is said that a photon
acquired a particular polarization (horizontal or vertical) only
after the photon has passed through a PBS. This process is random
and it is not determined by any physical reality.

The interpretation is significantly different within the approach
proposed here. Each photon is in a particular elementary state.
This elementary state describes a physical reality, which
predetermines the result of the photon interaction with PBS for
any orientation of the horizontal and vertical directions, that
is, in any polarization base. Thus, for any given orientation of
the polarization base it is predetermined in advance, which of the
two possible directions the photon will follow after passing
through the PBS. However, if we do not know in advance that the
photon has a certain polarization in a given base (belongs to a
beam having vertical or horizontal polarization), we cannot make
the corresponding prediction.

As a result of observing a photon passing through a PBS we acquire
certain information. Namely, we learn what the photon polarization
in the polarization base connected with the PBS was. In general,
the photon polarization changes uncontrollably when it passes
through a PBS. However, if a PBS performs a reproducible
measurement, then the polarization along the direction of the
polarization base associated with the PBS does not change. It does
change over the other directions. Thus, we can acquire the
information on the photon polarization along any single direction,
but only along one such direction.

The second device is a (simple) beamsplitter BS, which is used for
mixing two photon beams. Demonstrably one can imagine such a
device in the form of a semitrans-parent plate (see Fig. 5). Beams
of polarized photons fall on the two sides of this plate at the
same angle and in the same plane, perpendicular to the plate.

 \begin{figure}[h]
 \begin{center}

  \includegraphics{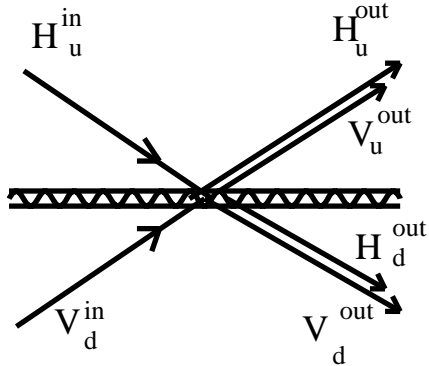}

   \caption{Simple beamsplitter.}
\end{center}
\end{figure}

In this case the device geometry also determines the polarization
base. We will assume that the horizontal direction lies in the
beam's plane, and the vertical one is perpendicular to this plane.

If the photons in two beams fall on the plate not simultaneously,
then each of them either passes through the plate, or is reflected
without changing the polarization. Both possibilities are entirely
random, and have equal probabilities. While if two photons from
different beams hit the plate simultaneously, then they interfere
according to the following rule

  \bea{23}
|H,V\ra_{u}^{in}&\to&\frac{1}{\sqrt{2}}\lt[|H,V\ra_u^{out}+
|H,V\ra_d^{out}\rt],\nn
|H,V\ra_{d}^{in}&\to&\frac{1}{\sqrt{2}}\lt[|H,V\ra_u^{out}-
|H,V\ra_d^{out}\rt].
 \eea
The symbol   $|H,V\ra$ in Eq.\rr{23}  denotes that the photon has
either horizontal polarization, that is, it is in the quantum
state  $|H\ra$, or the vertical one, that is, it is in the quantum
state | $|V\ra$. For every line the polarizations in the left and
right hand sides of Eq. \rr{23} coincide, while the polarizations
in different lines may be different. The index $u$($d$) denotes
that the photon belongs to the upper (lower) beam, while the index
$in(out)$ denotes that the photon belongs to the incident
(outgoing) beam.

An interesting result is obtained when the incident photons are in
one of the Bell states (given by Eq. \rr{19}). Using Eq. \rr{23},
we find that when passing through a simple beam splitter BS the
Bell states are transformed as follows.

  \bea{24}
|\Psi^{(-)}\ra_{ud}^{in}&\to&\frac{1}{\sqrt{2}}\lt[|H\ra_u^{out}|V\ra_d^{out}-
|V\ra_u^{out}|H\ra_d\rt]\equiv-|\Psi^{(-)}\ra_{ud}^{out},\nn
|\Psi^{(+)}\ra_{ud}^{in}&\to&\frac{1}{\sqrt{2}}\lt[|H\ra_u^{out}|V\ra_u^{out}-
|H\ra_d^{out}|V\ra_d^{out}\rt],\nn
|\Phi^{(-)}\ra_{ud}^{in}&\to&\frac{1}{\sqrt{2}}\lt[|H\ra_u^{out}|H\ra_u^{out}-
|V\ra_u|V\ra_u^{out}-|H\ra_d^{out}|H\ra_d^{out}+
|V\ra_d^{out}|V\ra_d^{out}\rt],\nn
|\Phi^{(+)}\ra_{ud}^{in}&\to&\frac{1}{\sqrt{2}}\lt[|H\ra_u^{out}|H\ra_u^{out}+
|V\ra_u^{out}|V\ra_u^{out}-|H\ra_d^{out}|H\ra_d^{out}-
|V\ra_d^{out}|V\ra_d^{out}\rt].
 \eea
These formulas imply that if the incident photon is in a singlet
state $|\Psi^{(-)}\ra_{ud}^{in}$, then the outgoing photons are on
different sides of the plate BS. While if the incident photon is
in one of the triplet Bell states
 $|\Psi^{(+)}\ra_{ud}^{in}$,
$|\Phi^{(-)}\ra_{ud}^{in}$, $|\Phi^{(+)}\ra_{ud}^{in}$, then both
outgoing photons are on the same side of the plate.

Equations \rr{23} and  \rr{24} give a concise and simple
mathematical description of the beam splitter's BS impact on beams
of polarized photons. However, below we will need a description of
the beam splitter's BS impact on isolated pairs of photons. In
order to describe individual photons mathematically in this case
we will have to deal with elementary states of these photons.
Then, we will have to use the equivalence classes of elementary
states for description of the quantum state of the photons.

There is a subtle distinction between the quantum states defined
as equivalence classes of elementary states, on the one hand, and
as Hilbert space vectors on the other hand. Although, both
definitions describe pure states (see \cc{slav1}). The point is,
that averaging the observables over different equivalence classes
$\Psi'=\{\vp\}_{\vp_{\eta'}}$ è $\Psi''=\{\vp\}_{\vp_{\eta''}}$
may sometimes yield identical results for the mean values of these
observables. That is, it may happen that $\Psi'(\A)=\Psi''(\A)$
holds for all observables $\A$.

Let us take a look what can be the consequences of that for a
physical system consisting of two particles with spin 1/2. The
algebra  $\AAA$ of spin observables in this system, as well as in
many others, has two maximal commutative subalgebras $\QQ'$ and
$\QQ''$. The generators of the subalgebra $\QQ'$ are the
observables  $\s_{1z}$, $\s_{2z}$, while the generators of the
subalgebra $\QQ''$ are the observables $\s_z=\s_{1z}+\s_{2z}$,
$\s^2$. Since $\s^2$ does not commute with  $\s_{1z}$ è $\s_{2z}$,

the subalgebras  $\QQ'$ and $\QQ''$ are different. However, they
have common elements. One such element is the one-dimensional
projector $\p=(1+2\s_{1z})(1+2\s_{2z})/4$. It is obvious that
$\p^*=\p$ è $\p^2=\p$, that is,  $\p$ is in fact a projector. The
commonly used representation of spin observables by operators in a
Hilbert space is exact, that is, different operators correspond to
different observables. In this representation a projection
operator on a one-dimensional Hilbert subspace corresponds to the
observable  $\p$. Therefore, the projector  $\p$ is
one-dimensional. Obviously the operator  $\p$ commutes with the
observables $\s_{1z}$,  $\s_{2z}$ $\s_{z}$. It is easy to verify
that  $\p$  commutes with  $\s^2$ as well. Therefore,  $\p\in
\QQ'\bigcap\QQ''$.

As was shown in \cc{slav1}, for each one dimensional projector
$\p$ belonging to some commutative subalgebra $\QQ$ the formula

  \beq{25}
\p\A\p=\Psi(\A)\p
 \eeq
defines a linear functional  $\Psi(\cdot)$ on the algebra  $\AAA$.

This functional describes mean values of observables in a pure
quantum state specified by particular values of a complete set of
commuting observables. Substituting the observables $\s_{1z}$,
$\s_{2z}$, $\s_{z}$, and  $\s^2$ instead  $\A$ in Eq.\rr{25} one
obtains
 \beq{26}
  \Psi(\s_{1z})=\Psi(\s_{2z})=\frac12,\quad\Psi(\s_z)=1, \quad
  \Psi(\s^2)=2.
 \eeq

Let us now form two equivalence classes  $\Psi'$ and  $\Psi''$.
The equivalence class  $\Psi'$ is specified by the values of the
observables $\s_{1z}$ and  $\s_{2z}$: $S_{1z}=S_{2z}=1/2$. This
means that all elementary states from the class  $\Psi'$, are
stable with respect to the observables from subalgebra  $\QQ'$,
and these observables obtain the specified values in these states.
Analogously, the class  $\Psi''$ is fixed by the values of the
observables  $\s_z$, $\s^2$: $S_z=1$ and  $S^2=2$. The elementary
states from the class  $\Psi''$ are stable with respect to the
observables from the subalgebra  $\QQ''$.

Equation \rr{26} implies that the functional  $\Psi(\cdot)$
defined by Eq. \rr{25} can play the role of statistical average,
both for the class  $\Psi'$ and for the class  $\Psi''$. If now we
use the functional  $\Psi(\cdot)$ and the canonical GNS
construction to build a representation of the algebra  $\AAA$ in
the Hilbert space  $\HHH$, then the equivalence classes  $\Psi'$
and $\Psi''$ would correspond to the same state vector in this
space.

On the other hand, if we define the quantum states  $\Psi'$ and
$\Psi''$ as the corresponding equivalence classes, then physically
such quantum states will be different. In the quantum state
$\Psi'$ each individual measurement must yield
$S_{1z}=S_{2z}=1/2$, while the values   $S^2=2$ and  $S_z=1$ are
obtained only in average, although the corresponding distributions
have zero variance. On the contrary, in the quantum state $\Psi''$
each individual measurement must yield  $S^2=2$ and  $S_z=1$,
while the values  $S_{1z}=S_{2z}=1/2$ are obtained only in
average, and again the corresponding distributions have zero
variance. The difference is very subtle; however, as we will see
below, it can manifest itself in the experiment.

Let us now go back to the discussion of the simple beam splitter
BS. We will consider it as a component of the classical measuring
device. In the general case the measuring device is composed of
two parts: an analyzer and a detector. Sometimes these parts can
be combined. An analyzer is a device with a single entrance
channel and a few exit ones. The investigated physical system
reaches the analyzer through the entrance channel. Depending on
the elementary state of this system the analyzer forwards it to
one of the exit channels. The detector registers through which
channel the system emerged from the analyzer. This is how
experimentalists obtain the information they are interested in
about the elementary state of the investigated system.

The beam splitter BS can be considered as an analyzer with two
exit channels. One channel (channel 0) corresponds to the photon
emission on different sides of the beam splitter's BS plate, the
other channel (channel 1) corresponds to the photon emission on
the same side of the plate.

As was already mentioned in Section 2, each measuring device is
calibrated using the values of observables from a certain maximal
commutative subalgebra  $\QQQ$. In the context of an analyzer this
means that each exit channel corresponds to certain values
(intervals of values) of observables from the subalgebra  $\QQQ$.
Therefore, the reaction of the analyzer is determined not by all
the characters comprising the elementary state of the investigated
system, but only by the character $\vx$. Accordingly, using the
analyzer one can obtain not the complete information about the
elementary state, but only the information about $\vx$.

For a photon, characteristics describing the polarization can
assume two values  $H$ and $V$. A certain character corresponds to
each polarization base (each orientation of BS). Thus, the
reaction of a concrete beam splitter BS is determined by the
characteristics related to the polarization base of this beam
splitter.

Let us consider the case when both photons arriving at the beam
splitter BS have the same polarization in the base associated with
this beam splitter, for instance, the vertical polarization. Then,
the initial quantum state of these photons can be described by the
vector $|V\ra^{in}_u|V\ra^{in}_d$. According to Eq. \rr{23} the
beam splitter transforms this vector as follows
 \beq{27}
|V\ra^{in}_u|V\ra^{in}_d\longrightarrow\frac12 \lt[
|V\ra^{out}_u|V\ra^{out}_u- |V\ra^{out}_d|V\ra^{out}_d\rt].
 \eeq

Vectors of the Hilbert space provide a statistical description of
a physical system. Therefore, Eq. \rr{27} itself maybe interpreted
as follows. If the incident photons had identical (vertical)
polarization in the beam splitter base, then after passing through
the beam splitter they will reach the exit channel with a
probability of 1.

Following our general logic we will assume that one can make a
stronger statement. Let the elementary states of the incident
photons be such that both photons have identical (vertical or
horizontal) polarization in the beam splitter base. Then, after
passing through the beam splitter these photons will reach the
exit channel 1. Here, we are not dealing with probabilities, but
we assume that this statement is valid for each pair of photons.

Using Eq. \rr{23} one can obtain the formula
 \beq{28}
|\vartheta\ra^{in}_u|\vartheta\ra^{in}_d\longrightarrow\frac12
\lt[ |\vartheta\ra^{out}_u|\vartheta\ra^{out}_u-
|\vartheta\ra^{out}_d|\vartheta\ra^{out}_d\rt],
 \eeq
analogous to Eq. \rr{27}, but it is also valid in the cases when
the vertical polarization is replaced by the polarization at some
angle $\vartheta$ ($\vartheta\neq n\pi/2$). In Eq. \rr{28}
$|\vartheta\ra=|V\ra\sin\vartheta + |H\ra\cos\vartheta$.

The stronger statement made in the previous paragraph is no longer
valid in this case. Indeed, in the language of elementary states
the fact that both photons have identical polarization along the
direction corresponding to the angle $\vartheta$ means the
following. These two photons have identical characteristics
$\vp_{\vartheta}$ describing the polarization along the direction
corresponding to the angle $\vartheta$. However, in the general
case the characteristics  $\vp_{n\pi/2}$ corresponding to the same
elementary state will be different for these photons. The opposite
would mean that we succeeded in creating a quantum state having
definite polarizations along two directions, although the
corresponding observables are incompatible. However, according to
the statement made earlier, it is the characteristics
$\vp_{n\pi/2}$ that determine the behavior of the outgoing
photons. Therefore, only the probabilistic interpretation of
Eq.\rr{28} is possible in this case. Such interpretation is in
agreement with Postulate 6.

Now we are ready to discuss the real experiment \cc{bpw}. The
layout of the experimental apparatus is shown in Fig. 6.

\begin{figure}[h]
 \begin{center}

  \includegraphics{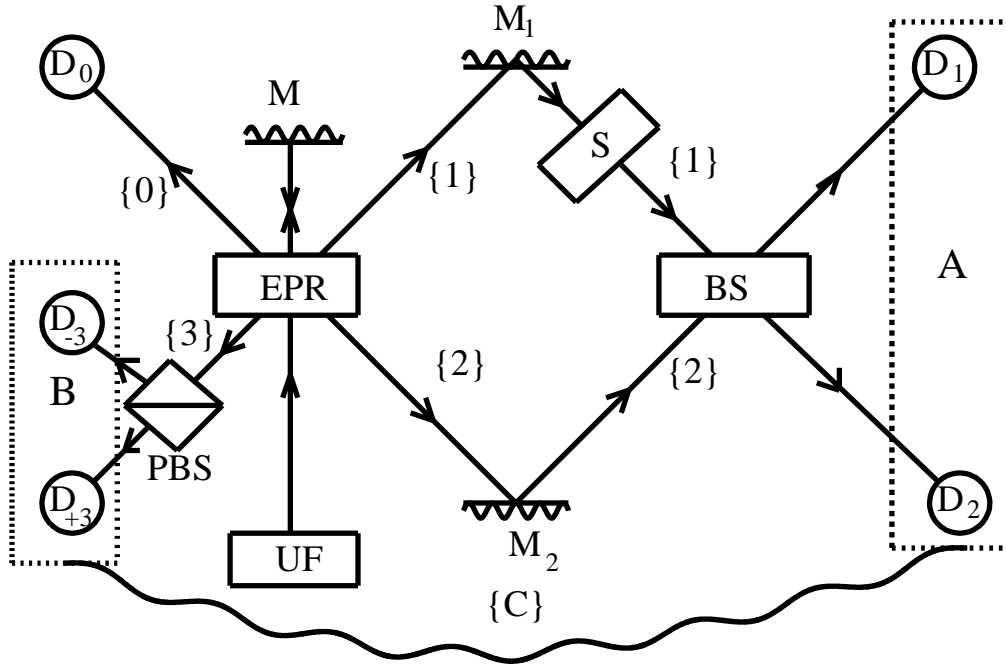}

   \caption{The layout of the quantum teleportation experiment.}
\end{center}
\end{figure}

Here, $UF$ is the ultraviolet laser (a source of photons); $EPR$
is the source of EPR pairs;  $M$, $M_1$, and  $M_2$ are mirrors;
the mirror  $M$ is adjustable;  $BS$ is a simple beam splitter;
$PBS$ is a polarization beam splitter;  $S$ is the encoder of the
initial state of photon  \{1\}; $D_0$, $D_1$, $D_2$, $D_{+3}$, and
$D_{-3}$ are single-particle (reacting on a single photon)
detectors; and \{C\} is the coincidence circuit for the signals
from detectors (a classical communication channel).

The detectors  $D_1$ and  $D_2$ belong to the  $A$ (Alice) zone,
the detectors $D_{+3}$ and  $D_{-3}$ belong to zone  $B$ (Bob). To
simplify the discussion we will assume that all the photons
propagate in the plane of the picture. The horizontal direction of
the beam splitter's BS polarization base and the horizontal
direction of the initial polarization base of the beam splitter
PBS he in the same plane. The base of the PBS beam splitter
changes in the process of the experiment.

The apparatus works as follows. The ultraviolet laser  $UF$ emits
a photon pulse. This pulse reaches the nonlinear crystal EPR,
where an EPR pair consisting of photons \{0\}-\{1\} is created.
The pulse goes through EPR, hits the mirror  $M$, is reflected by
the mirror, and reaches the EPR again, where the second EPR pair
consisting of photons  \{2\}-\{3\}.

The photon  \{0\} plays an auxiliary role and it is immediately
forwarded in the detector  $D_0$. Its partner in the EPR pair, the
photon \{1\}, is reflected by the mirror  $M_1$, and hits the
encoder  $S$, where it acquires a certain polarization. After that
the photon  \{1\} reaches the beam splitter  $BS$. The photon
\{2\} belonging to the second EPR pair is reflected by the mirror
$M_2$ and reaches the same beam splitter  $BS$, but from the other
side.

By adjusting the mirror  $M$ one can create the following two
regimes of the $BS$ operation: either the photons  \{1\} and \{2\}
reach the beam splitter  $BS$ simultaneously, or they reach it at
different times. After the beam splitter  $BS$ the scattered
photons hit the detectors  $D_1$ and  $D_2$. Here, the following
three alternatives are possible: either both photons reach the
detector $D_1$, or both reach the detector  $D_2$, or one of them
reaches detector  $D_1$ and the other one detector $D_2$.

The photon  \{2\} partner in the EPR-pair --- photon  \{3\} is
forwarded in the polarization beam splitter PBS. If photon \{3\}
has the horizontal polarization in the base of this beam splitter,
then having left the beam splitter it is forwarded to the detector
$D_{+3}$, if it has the vertical polarization, then it is
forwarded to the detector  $D_{-3}$.

The coincidence circuit  \{C\} is tuned in such a way, that it
registers the events of simultaneous triggering of either the
detectors $D_0$, $D_1$, $D_2$, and $D_{+3}$, or the detectors
$D_0$, $D_1$, $D_2$, and  $D_{-3}$. The presence of the signal
from  $D_0$ indicates that photon \{1\}, the partner of photon
\{0\}, takes place in the registered process. The state of photon
\{1\} was subjected to encoding, which was necessary for excluding
accidental coincidence of the signals from $D_1$, $D_2$, and
$D_{+3}$; or from the detectors  $D_1$, $D_2$, and $D_{-3}$. The
source of such accidental registrations can be, for instance, the
event where two photons  \{2\} almost simultaneously reach the
beam splitter BS.

The authors of the experiment interpreted a simultaneous
triggering of the detectors  $D_1$ and  $D_2$ as observing photons
\{1\} and  \{2\} in the state  $\Psi^{(-)}_{12}$. According to the
standard description of the teleportation phenomenon, in this
case, photon \{3\} must acquire the same polarization as photon
\{1\} after passing through the encoder $S$. In order to check if
this is indeed the case, the polarization beam splitter  $PBS$ was
oriented in such a way that if the photon  \{3\} polarization
coincides with the photon  \{1\} polarization, then after passing
through the  $PBS$  ôîòîí \{3\} must reach the detector $D_{+3}$.
Therefore, the coincidence of the signals from the detectors
$D_0-D_1-D_2-D_{+3}$ and the absence of coincidence
$D_0-D_1-D_2-D_{-3}$ corresponds to the presence of teleportation.

In fact, the frequency of this type of coincidences depending on
the location of the mirror  $M$ was measured in the experiment.
When the location was such that the photons \{1\} and  \{2\} reach
the  $BS$ beam splitter simultaneously, the frequency of the
$D_0-D_1-D_2-D_{-3}$ coincidences must drop to zero. Otherwise the
numbers of  $D_0-D_1-D_2-D_{+3}$ and  $D_0-D_1-D_2-D_{-3}$
coincidences must become identical. Indeed, the graph of the
number of  $D_0-D_1-D_2-D_{-3}$ coincidences as a function of the
mirror  $M$ location shows a clearly visible drop, approximately
by one order of magnitude. This fact was considered as proof that
teleportation was taking place.

It is very difficult to explain this phenomenon from the physical
point of view. Indeed, initially the photons  \{1\} and  \{3\}
were independent. In the subsequent manipulations connected with
the encoder $S$, beam splitter  BS and detectors  $D_1$ and $D_2$,
the photon  \{3\} did not take part at all. It is absolutely
unclear how it could acquire the same polarization as that of
photon  \{1\} after coding.

Even more striking results were obtained in the subsequent
experiments  \cc{jwpz}. In the whole, the results of these
experiments confirmed the results of the previous ones, but
another diflficult-to-explain phenomenon was observed. By changing
the lengths of the paths traveled by different photons the
following was achieved in the experiment reported in \cc{jwpz}.
The presence of teleportation (arrival of the signal from the
detector  $D_{+3}$ and absence of the signal from  $D_{-3}$) was
established even before the conditions for this teleportation were
created (triggering of the detectors  $D_1$ and  $D_2$). We would
like to remind the reader that, according to the standard version,
teleportation is a consequence of the measurements performed by
the detectors  $D_1$ and  $D_2$. Moreover, in principle, one could
create a situation when the signals from the detectors $D_3$ and
$D_{-3}$ arrive before the photon  \{1\} reaches the encoder  $S$.
That is, photon  \{3\} is able to foresee, which polarization will
be acquired by photon  \{1\}. Thus, in this case we are witnessing
not only a spatial nonlocality but also a temporal one.

Strictly speaking all the above does not contradict the
conclusions of the standard quantum mechanics, because the quantum
state there (in the Heisenberg picture) provides the information
on the behavior of physical systems at any instant of time.
However, from the conventional logic point of view all this looks
like complete absurdity. At the same time, in the framework of the
approach used in this paper the results of the quantum
teleportation experiment can be easily put in agreement with the
conclusions of the standard (classical) formal logic.

In reality, the quantum teleportation phenomenon was investigated
in the experiments  \cc{bpw} for two cases of the photon \{1\}: at
$90^0$ and  $45^0$ in the initial base.

Let us begin with the case when photon  \{1\} is polarized at the
angle of  $90^0$. In this case the polarization beam splitter
$PBS$ is also rotated by  $90^0$. Accordingly, if photon  \{3\} is
polarized vertically in the initial base, then the signal will
arrive from the detector  $D_{+3}$, while if it is polarized
horizontally, then from the detector  $D_{-3}$. Let the signal
have arrived from the detector  $D_{-3}$. This means that the
photon \{2\} making up an EPR pair with the photon  \{3\}, is
polarized vertically. In this case the photons  \{1\} and  \{2\},
arriving at the beam splitter  $BS$, will have the identical
polarization and will reach the exit channel~1. That is, both
photons reach the same detector: either detector  $D_1$, or
detector  $D_2$. As a result, the coincidence circuit will
register that the  $D_0-D_1-D_2-D_{-3}$ coincidence is  absent.
Note that it is absolutely irrelevant which of the photons
\{0\},\{1\},\{2\}, or \{3\} reach the designated detector first.
What is important is that the photons  \{1\} and  \{2\} reach the
beamsplitter BS simultaneously. Of course, the coincidence circuit
must take into account the time necessary for each photon to reach
the corresponding detector. All this is quite consistent with the
classical logic, and does not require any special quantum
nonlocality of the measuring procedure, neither in time, nor in
space.

Let us now consider the case when the encoder  $S$ polarizes the
photon  \{1\} at the angle of  $45^0$ in the initial base.
Accordingly, we will assume that the polarization beam splitter
PBS is also rotated by  $-45^0$ relative to the initial base.
Repeating the previous argument we conclude that photons \{1\} and
\{2\} arriving at the BS beam splitter will have identical
polarizations at the angle  $45^0$ in the beam splitter's base.
Therefore, it seems that, as in the previous case, they must reach
the exit channel 1 of the BS beam splitter. However, the
beamsplitter BS classifies the outgoing photons to different
channels depending on whether they have {\it horizontal} or  {\it
vertical} polarization in the BS base. Various combinations of
horizontal and vertical polarizations are possible for each pair
of photons polarized at  $45^0$. Therefore, having passed through
the BS beam splitter the photons \{1\} and  \{2\} may in some
cases reach both the exit channel 1 and the exit channel 0.

Therefore, the $D_0-D_1-D_2-D_{-3}$ coincidences are possible.
That is, the teleportation will not be observed in these cases.

A special parameter --- the teleportation fidelity --- is used for
description of quantum teleportation. This parameter characterizes
the relative number of successful teleportation events. Therefore,
the teleportation fidelity in the case of  $45^0$ polarization
must be lower than in the case of $90^0$ polarization. However,
even in the former case the teleportation fidelity should be close
to 1.

Indeed, Eq. \rr{27} and the corresponding probabilistic
interpretation can be applied in the case of  $45^0$ polarization.
This means that with a probability of 1 the photons emerging from
the BS beam splitter must reach the exit channel 1. Therefore,
under ideal conditions the average value of the teleportation
fidelity must tend to one with an increasing number of trials.

However, in real experiments the number of cases when the presence
of  $D_0-D_1-D_2-D_{+3}$ coincidences and the absence of
$D_0-D_1-D_2-D_{-3}$ coincidences, or vice versa, registered is very
 small. Therefore, the teleportation fidelity registered in these
 events can be noticeably different from the theoretical mean
 value.

It is worth mentioning that, as was established in the experiments
reported in  \cc{jwpz}, the teleportation fidelity reaches its
maximum for the polarization angles  $0^0$ and  $90^0$, while for
the $45^0$ the fidelity is minimal. This is in agreement with the
above discussions. However, the authors of the experiment are
inclined to interpret this fact as a measurement error. It would
be very desirable to investigate this effect experimentally in
greater detail.

As the statistics improve, the difference between the
teleportation fidelity for $90^0$ and  $45^0$ must decrease.
However, this does not mean that an improvement in statistics does
not help us discover in this phenomenon deviations from
predictions of the standard quantum mechanics. But, we should
treat these deviations not as measurement errors, but as one of
the investigated effects. For instance, one can study the quantity
   $$     \rho=\frac{N_-(45)}{N_-(90)} \qquad \mbox{ïðè}\quad
   N_-(45)+N_+(45)=N_-(90)+N_+(90),   $$
where $N_-(45)$ is the number of  $D_0-D_1-D_2-D_{-3}$
coincidences in the absence of $D_0-D_1-D_2-D_{+3}$ coincidences,
when photon \{1\} is polarized at $45^0$,  $N_+(45)$ is the number
of $D_0-D_1-D_2-D_{+3}$ coincidences in the absence of
$D_0-D_1-D_2-D_{-3}$, while  $N_-(90)$ and  $N_+(90)$ are the same
quantities for the polarization $90^0$.

According to the rules of the standard quantum mechanics the
quantity  $\rho$ must tend to one as statistics improves. In the
framework of the approach proposed in this paper the quantity
$\rho$ will always remain greater than one.

\section{CONCLUSION}

Summarizing, we would like to recall the words of the remarkable
physicist Richard Feynman~\cc{feyn2}.

"On the other hand, I think I can safely say that nobody
understands quantum mechanics. So do not take the lecture too
seriously, feeling that you really have to understand in terms of
some model what I am going to describe, but just relax and enjoy
it.

Do not keep saying to yourself, if you can possibly avoid it, 'But
how can it be like that?' because you will go 'down the drain',
into a blind alley from which nobody has yet escaped. Nobody knows
how it can be like that."

Feynman himself did not always follow his own "pia desideria." The
present paper also goes against his advices. Here, we present a
model based on the novel for the quantum mechanics notion of the
"elementary state." This notion is an attribute of individual
physical systems, and it is introduced in order to eliminate the
incompleteness of the quantum mechanics pointed out by Einstein.

Unlike the traditionally used model of quantum mechanics, the
model presented here does not have sharp contradictions with
classical concepts. This allows one to create a more or less
intuitively clear picture of the quantum world. In particular, our
model allows one to present an intuitively appealing
interpretation of quantum phenomena, whose traditional
interpretation looks absolutely absurd from the classical physics
point of view. The list of such phenomena includes quantum
particle scattering on a double-slit screen, the
Einstein-Podolsky-Rosen paradox, the delayed choice experiment,
and quantum teleportation.

Within the traditional approach interpretations of all these
phenomena relied on the idea of nonlocality supposedly inherent in
quantum measurements. In the framework of the model proposed here
it was possible to show that these phenomena contradict neither
the causality principle, nor the locality principle in its
strictest form. Namely, it was shown that these phenomena agree
with the idea of the existence of a local physical reality, which
determines the behavior of a physical system in a bounded domain
of the Minkowski space.

At the same time the considered model does not contradict the
traditional mathematical formalism of the quantum mechanics. It
only supplements the traditional formalism and specifies the
corresponding boundaries of the applicability domain.

\end{document}